\documentclass[a4paper,12pt]{article}
\usepackage{jheppub}

\usepackage{amsmath,amssymb,multirow,array}

\title{{ The Energy Loss of a Heavy Quark Moving Through a General Fluid Dynamical Flow}}

\author{Navid Abbasi$^{a,b}$\footnote{Abbasi@ipm.ir}, \ Ali Davody$^b$\footnote{Davody@ipm.ir},\\
\small{\emph{$^{a}$Department of Physics, Sharif University of Technology,}} \\
\small{\emph{P.O. Box 11365-9161, Tehran, Iran}} \\ [1mm]
\small{\emph{$^{b}$ School of Particles and Accelerators, Institute for Research in Fundamental Sciences (IPM), P.O. Box 19395-5531, Tehran, Iran}} 
}

\abstract{
We determine the most general form of the covariant drag  force exerted on a quark moving through  a fluid dynamical flow.
 Up to first order in derivative expansion, our general formula requires the specification of seven
coefficient functions. We use the perturbative method introduced in arXiv:1202.2737 and find all these coefficients
in the hydrodynamic regime of a  $\mathcal{N}=4$ SYM plasma. Having this general formula, we can obtain the rate of the energy
and momentum loss of a quark, namely the drag force, in a general flow. This result makes it possible to perturbatively study the motion of heavy quarks
moving through the Bjorken flow up to first order in derivative expansion.
}

\begin{document}

\setlength{\baselineskip}{16pt}
\begin{titlepage}
\maketitle

\vspace{-36pt}

\thispagestyle{empty}
\setcounter{page}{0}
\end{titlepage}

\renewcommand{\baselinestretch}{1}  
\tableofcontents
\renewcommand{\baselinestretch}{1.2}  

\section{Introduction}
\label{intro}
The only thing that we obtain from heavy ion collisions experiment is the  spectrum of the hadrons coming out 
of a fireball.  As it is well known, after such collision, a new phase of system is produced  which
is the most strongly coupled matter known in nature \cite{Shuryak:2003xe}. 
Having the spectrum of hadrons, we may phenomenologically find several properties of the initial state and evolution of this hot and
dense matter. Although the right place to explore these phenomena  is  QCD, due to failure of  perturbative methods in
the strong coupling regime, progress looks formidable.

 As a well known replaced tool, the  AdS/CFT duality \cite{Maldacena:1997re,Witten:1998qj} has been trying to investigate some
aspects of the problem during recent years. In one of the preliminary attempts in this direction, \cite{Herzog:2006gh} and
\cite{Gubser:2006bz} computed  the drag force exerted on a  heavy quark moving through a $\mathcal{N}=4$ SYM thermal plasma.
 Their computations are based on studying of a classical string dynamics in a five dimensional  AdS black brane
dual to the four dimensional  boundary thermal plasma through which the quark is moving. 
Based on their result, Horowitz and Gyulassy showed that the AdS/CFT duality predicts a decreasing nuclear modification factor of bottom jets,
 $R^{AdS}_{AA}(p_{T} )$, while the  pQCD energy loss gives an increasing $R^{PQCD}_{AA}(p_{T} )$ \cite{Horowitz:2007su}.

On the other hand, there are some evidence confirming that  just shortly after the scattering, the evolution of  the QGP is  described
 with hydrodynamic approximation (See \cite{Kolb:2003dz} and references therein.). So  to theoretically produce a
more precise hadron spectrum compatible with the one obtained from data, it seems necessary to take the hydrodynamical
evolution of the QGP into account. As it is well known, a relativistic fluid dynamical flow consistent with the final detected
particles in the QGP experiment is the Bjorken flow \cite{Bjorken:1982qr}. So our interesting problem would be computing the rate of the energy loss
of a heavy quark moving through such flow. To this end, we must study  the dynamics of a classical string in the gravity background dual 
to a boundary Bjorken flow\footnote{It should be noted that in  \cite{Horowitz:2007su}, the authors have used the AdS/CFT drag force replacing the 
temperature with the zero order temperature of the Bjorken flow therein.}.

Relatedly, the Fluid/Gravity  correspondence \cite{Bhattacharyya:2008jc}  provides our desired  dual background. Fluid/Gravity duality
is able to   perturbatively constructs the asymptotically  AdS5 background dual to every boundary fluid  dynamical flow, order by order,  in a boundary
derivative expansion. In \cite{Abbasi:2012qz} we explained "how" to use this duality and to compute the drag force in a general flow perturbatively .
Then as  an example we computed the first order corrected drag force exerted on  specific quarks in Bjorken flow, i.e. the quarks constrained 
to move in the zero rapidity plane. More importantly, it should be noted that our method in \cite{Abbasi:2012qz} was not able to give the drag force exerted
on a quark with non-vanishing initial rapidity. In this paper however, we give a solution for the "covariant drag force" in a general fluid dynamical flow,
completely answering the questions remained unresolved about a general quark moving through The Bjorken flow.

Before proceeding further, let us exactly explain what we mean by  the "drag force" and "covariant drag force". 
When an external  quark moves through the medium, the energy-momentum tensor of the medium is not 
conserved as before, instead we have \cite{Chesler:2007sv}:
\begin{equation}
\partial_{\nu} T^{\mu \nu}(x)=\;- f^{\mu} \delta^3(\overrightarrow{x}-\overrightarrow{x_{0}}(t)).
\end{equation}
As expected, the right hand side determines  the volume density of the energy and momentum
pumped from the quark into the medium per unit time. \footnote{$\overrightarrow{x}=\overrightarrow{x}_{0}(t)$
is the quark's path in the medium.} $f^{\mu}$ is the \textbf{drag force} and gives the  rate of the energy and
momentum loss of the quark. It is clear that the drag force does not transform under Lorentz transformations as 
a four vector. On the other hand, what we will refer to as the \textbf{covariant drag force} is:
\begin{equation}
F^{\mu}=\; \frac{dp^{\mu}}{d\tau}=\; \gamma  \frac{dp^{\mu}}{dt}=\; \gamma f^{\mu}=\; \gamma \; (\;\frac{dE}{dt},\frac{d\overrightarrow{p}}{dt}).
\end{equation} 

What we want to do in this paper is to compute $F^{\mu}$ covariantly  in a general fluid dynamical flow and then to find  the drag force $f^{\mu}$.
To proceed, we first determine the most general  form of the covariant drag force up to first order in derivative expansion, independent of 
AdS/CFT methods. This general four vector is given in terms of the quark velocity, local fluid dynamical variables and their derivatives.
 At zeroth order, namely ideal fluid,  there is only one coefficient function which has to be specified. In $\mathcal{N}=4$ SYM field theory, this coefficient has been already found  \cite{Abbasi:2012qz}\footnote{$\tilde{u}^{\mu}$ stands for the quark velocity. $u^{\mu}(x)$ and $b(x)=\frac{1}{\pi T(x)}$ define a 
dynamical flow compatible with fluid dynamics equations. In addition, $x$ refers to all space-time points.}. Rigorously speaking, covariant
drag force to this order may be given by:
\begin{equation}\label{1}
F^{\mu}_{ideal}(x)=\frac{1}{2\pi \alpha'} \frac{1}{b(x)^2}((\tilde{u} . u(x)) \tilde{u}^{\mu}+u^{\mu}(x)).
\end{equation}
At first order however, this formula acquires  derivative corrections  constructed from fluid dynamical variables.\footnote{Recall
that in\cite{Herzog:2006gh},\cite{Gubser:2006bz} and \cite{Abbasi:2012qz} and this work, it  has been assumed that quark
is dragging with constant velocity through the medium.} Formally, the most general corrections may appear as the following:
\begin{equation}\label{2}
F^{\mu}(x)=(1+\sum_{i} \alpha_{i} s_{i})\; F^{\mu}_{ideal}(x)+\sum_{i} \beta_{i} v_{i}^{\mu}
\end{equation}
where $s_{i}$($v_{i}^{\mu}$) is a basis of on shell inequivalent one derivative scalars(vectors) and $\alpha_{i}$ and $\beta_{i}$
are unknown coefficient functions.
As it will be discussed in detail throughout the next section, there are three (three) inequivalent one derivative scalars (vectors) 
respecting symmetry considerations. We use the method we developed in \cite{Abbasi:2012qz}  to specify these six coefficient functions. We consider a quark moving
through a general fluid flow and compute the drag force  in its rest frame (RF) at one instant of time. Comparing the result 
 with the covariant formula evaluated in the same situation, we find  six equations  giving exactly these six coefficient functions.
 We therefore fully compute the covariant drag force in a general fluid flow of a $\mathcal{N}=4$ SYM field theory, up to first order in derivative expansion.
 
It is interesting to note that all  $\beta_{i}$ functions turn out to be zero, meaning that the vectorial structure of the drag force
remains unchanged  up to first order and just the scalar factor is corrected\footnote{Notice that at zero order the there exists just one
independent vector term.}.  This result is not predictable in the filed theory side and is specifically obtained in the $\mathcal{N}=4$ SYM field theory
field theory via holographic computations.
 
Considering the Bjorken flow as the relevant fluid flow to the QGP experiments,  we successfully 
reproduce the results of \cite{Abbasi:2012qz} by use of our general covariant formula.\footnote{In \cite{Abbasi:2012qz}, the corrections have been
computed through a different manner.}
 In addition, having obtained the covariant drag force, one can simply find the drag force exerted on a general
heavy quark  with arbitrary initial conditions in the Bjorken flow; what we were not able to obtain in \cite{Abbasi:2012qz}.

The paper is organized as it follows; in section \eqref{most} we determine the general form of covariant drag force up to fist order
in derivative expansion. In section \eqref{thermal} we first briefly review how to compute the drag force via AdS/CFT and then apply it to
the case of a globally boosted thermal plasma. In section \eqref{fluid} we first glimpse at the Fluid/Gravity duality then  we outline in detail
the structure of our perturbative computations. Finally we will compute different components of covariant drag force at one 
instant of time. In section \eqref{matching}, we proceed to exploit the results of previous sections to fix the unknown coefficients 
in the general covariant drag force. In addition we argue on the importance of such covariant formula from the view point of
experiment.  In section \eqref{open} we end with pointing the open questions.

\section{The most general form of the covariant drag force}
\label{most}
Let us recall  the relativistic constraint on the proper force exerted on a particle
\footnote{This constraint was truly met at ideal order in \eqref{1}.}
\begin{equation}\label{3}
\tilde{u}_{\mu}\;F^{\mu}=0.
\end{equation}
This condition forces the covariant drag force to belong to the  vector representation of the $SO(3)$ group 
orthogonal to the quark velocity $\tilde{u}$\footnote{This idea is reminiscent of the analogue idea used in preparing 
the first order energy-momentum tensor of relativistic fluid dynamics in Landau frame\cite{Bhattacharya:2011tra}. }.
In Table 1 we have listed all scalars and vectors
constructed by $u^{\mu}(x)$, $b(x)$ and $\tilde{u}^{\mu}$ compatible with \eqref{3}.
As it is clearly seen,  there is only one appropriate vector term at ideal order. So the covariant drag force at ideal order must
be as the following:
\begin{equation}\label{4}
F^{\mu}_{ideal}(x)\;=\;\xi(S_{1},S_{2})\;\tilde{p}^{\mu  \nu} u_{\nu}(x)
\end{equation}
with $\xi(S_{1},S_{2})$ the unknown scalar coefficient function. 

\begin{table}
\label{table one}
\centering
\begin{tabular}[h]{|c|c|c|}
\hline
\hline
Type & Data & Evaluated in RF \\
\hline
\hline
Scalars &$S_{1}=\tilde{u} . u(x)$ & $u_{0}(x)$ \\ 
              &$S_{2}= b(x)$ & $b(x)$  \\
             
\hline             
Vectors & $V_{1}^{\mu}=\tilde{p}^{\mu \nu}u_{\mu}(x)$  & $u^{\mu}(x)$ \\
 \hline
 \hline
\end{tabular}
\caption{Zero order data. $SO(3)$ scalars and vectors. Note that $\tilde{p}^{\mu \nu}=\tilde{u}^{\mu} \tilde{u}^{\nu} +\eta^{\mu \nu}$}
is the projection tensor on the three dimensional space-time orthogonal to $\tilde{u}^{\mu}$.
\label{tdpvfd}
\end{table}

In table Table 2 we have organized one derivative data. There exist both scalar and vector derivative terms contributing to 
first order. Scalar one derivative terms can contribute to the scalar factor of $F^{\mu}_{ideal}(x)$, while
every  vector one derivative term may  separately contribute to dreg force as an additive term. So to first
order in derivative correction, the most general covariant drag force is constructed by adding a linear combination of $s_{i}$ to
$\xi(S_{1},S_{2})$ accompanied by adding a linear combination of $v_{i}^{\mu}$ to  $\xi(S_{1},S_{2})$-corrected drag force as
the following:
\begin{equation}\label{6}
F^{\mu}(x)=\left(1+\sum_{i=1}^{3}  \alpha_{i}(S_{1},S_{2})\; s_{i} \right) F_{ideal}^{\mu}(x)+\;\sum_{i=1}^{3} \beta_{i}(S_{1},S_{2})\; v_{i}^{\mu} .
\end{equation}
Our goal is to determine the six coefficient functions, $\{\alpha_{i}(S_{1},S_{2})\}$ and $\{\beta_{i}(S_{1},S_{2})\}$.
It may be interesting to note that  the  condition \eqref{3} is automatically satisfied once we use the  $SO(3)$ decomposition.
It is because  both $F_{ideal}^{\mu}(x)$ and $v_{i}^{\mu}$ are transverse to $\tilde{u}^{\mu}$,
by construction.

\begin{table}
\renewcommand\arraystretch{1.1}
\centering
\begin{tabular}{| p{2.4 cm} | c | c | l |}
\hline  
\hline      
	 \raggedright $SO(3)$ classification & All  data & Equations of motion & Independent  data\\
\hline
		\hline
	\multirow{5}{*}{Scalars} & 
	$ \tilde{u}^{\mu} \tilde{u}^{\nu} \partial_{\mu} u_{\nu}$  & 
	\multirow{5}{*}{
		\begin{minipage}{3 cm}
			\begin{flushleft} 
				$\tilde{u}_{\mu} \partial_{\nu} T^{\mu \nu} =0$ \\ 
                                $u_{\mu} \partial_{\nu} T^{\mu \nu} =0$ 
			\end{flushleft}
		\end{minipage} 
	} &
	\multirow{5}{*}{
		\begin{minipage}{3 cm}
			\begin{flushleft} 
				$s_1 =  \tilde{u}^{\mu} \tilde{u}^{\nu} \partial_{\mu} u_{\nu}$ \\ 
                                $s_2 =\tilde{u}^{\mu}\partial_{\mu}b/b$ \\  $s_3 = u^{\mu} \tilde{u}^{\nu} \partial_{\mu} u_{\nu}$ 
			\end{flushleft}
		\end{minipage} 
	} \\
	&  $\tilde{u}^{\mu}\partial_{\mu}b$ & & \\
	&$ u^{\mu} \tilde{u}^{\nu} \partial_{\mu} u_{\nu}$ & & \\
	&$\partial_{\mu}u^{\mu}$ & &\\
	&  $u^{\mu}\partial_{\mu}b$ & & \\

\hline
	\multirow{4}{*}{Vectors} & 
	$\tilde{p}^{\mu \alpha} \tilde{u}^{\beta} \partial_{\alpha} u_{\beta }$ & 
	\multirow{4}{*}{$\tilde{P}^{\mu}_{\phantom{\mu}\nu}\partial_{\alpha}T^{\alpha \nu} =0$} &
	\multirow{4}{*}{
		\begin{minipage}{3 cm}
			\begin{flushleft} 
				$v_1^{\mu} =\tilde{p}^{\mu \alpha} \tilde{u}^{\beta} \partial_{\alpha} u_{\beta }$ \\ 
                                $v_2^{\mu} =\tilde{p}^{\mu \alpha} \tilde{u}^{\beta} \partial_{\beta} u_{\alpha }$ \\
                                $v_3^{\mu} =\tilde{p}^{\mu \alpha} u^{\beta} \partial_{\beta} u_{\alpha }$
			\end{flushleft}
		\end{minipage} 
	} \\
	& $\tilde{p}^{\mu \alpha} \tilde{u}^{\beta} \partial_{\beta} u_{\alpha }$ & & \\
	&$\tilde{p}^{\mu \alpha} u^{\beta} \partial_{\beta} u_{\alpha }$  & & \\
	& $\tilde{p}^{\mu \alpha}  \partial_{\alpha}b$ & & \\
\hline
\hline
	\end{tabular}
	\caption{One derivative data.}
\end{table}

Before ending this section let us explain how we have arranged derivative terms in different sectors of $SO(3)$ group
in Table 2. Recall that the basic one derivative data are $\partial_{\mu}b$  and $\partial_{\mu}u_{\nu}$. Derivative terms
in the scalar sector may be constructed by contracting these basic data with each of $u^{\mu}$, $\tilde{u}^{\mu}$ or $\eta^{\mu \nu}$.
The reason is that scalar terms are supposed to multiply by $F^{\mu}_{ideal}(x)$ which has been originally constructed compatible
with \eqref{3}. In the case of vector terms however, we must contract $\tilde{p}^{\mu \nu}$ with one-free-index derivative data.


\section{Drag force in a thermal state of $\mathcal{N}=4$ SYM plasma}
\label{thermal}
The starting point of the computations in hydrodynamic regime is to study the problem in a thermal state.
In this section we will investigate the motion of a quark in a strongly coupled thermal field theory through
studying its dual gravity picture.  The problem in gravity side is to study the dynamics of a classical string  with a specific boundary
condition in an asymptotically AdS5 background.
 Like \cite{Herzog:2006gh} and \cite{Gubser:2006bz}, we demand one end point of the string is to be constrained to move with a constant velocity on the boundary.
 It is  also required that the string to be trailed  into the bulk all the way from boundary to the horizon.

\subsection{Dynamics of a classical string and drag force}
\label{dynamics}
 In order to study the dynamics of  string, we have to solve the equations of motion (EoM) related to
 string's degrees of freedom.  Consider  the Nambu-Goto action:
 \begin{equation}\label{8}
S=- \frac{1}{2\pi \alpha'}\int d\tau d\sigma \mathcal{L} =-\frac{1}{2\pi \alpha'}\int d\tau d\sigma \sqrt{-g}\;\;\;\;\; \sigma \in [0,\sigma_{1}]
\end{equation}
with  $g=det g_{\alpha \beta}$ and $g_{\alpha \beta}=G_{MN}\partial _{\alpha} x^{M} \partial_{\beta} x^{N}$ the induced metric on the world-sheet.
\footnote{Latin indices $\{M,N, \cdots \}$ have been used to denote bulk directions $(r,t,\overrightarrow{x})$, while $\alpha$ and $\beta$ indices  refer to world-sheet coordinates $(\tau, \sigma)$.}
 We choose  $\tau=t$ and $\sigma=r$ to fix the reparametrization freedom on the world-sheet. So the most general 
embedding of the string  in the mentioned  background may be given in the following form:
\begin{equation}\label{9}
x^{M}_{sol}(r,t)=(r,\;t,\;X_{x}(r,t),\;X_{y}(r,t),\;X_{z}(r,t))
\end{equation}
where $X_{i}(r,t)$, $i=x,y,z$ are the string's degrees of freedom.  Throughout this paper, we always prefer to write
 EoM in the form of world-sheet currents conservation law:
 \begin{equation}\label{10}
\partial_{\alpha} \Pi^{\alpha}_{M}= f_{M}\;\;\;\;\;\;\;\;\; \alpha= t,r
\end{equation}
where
\begin{equation}\label{10'}
\Pi_{M}^{\alpha}=\frac{\partial{\cal{L}}}{\partial({\partial_{\alpha}x^{M}})}=-\frac{1}{2\pi\alpha'}\;\;\sqrt{-g}\;\;P_{M}^{\alpha}\,, \;\;\;\;\;\;\;\;\;\;\;
  f_{M}=\frac{\partial \mathcal{L} }{ \partial x^{M}}\,.
\end{equation}  
Let us quickly recall that in our picture, the end point of the string on the boundary is identified with the quark in the plasma.
So in order to compute the drag force we have to compute the momentum current  going down  into the bulk at this end point.  It would 
be nothing except the ingoing momentum flux crossing the time-like boundary of  world-sheet, $\sigma=\sigma_{1}$ path.
Assuming  $n^{\alpha}$ as the inward normal vector to this path, it turns out that:
\begin{equation}\label{11}
f^{\mu}=\;(\frac{dE}{dt},\frac{d\overrightarrow{p}}{dt})=\;- \Pi^{\mu}_{\alpha}\; n^{\alpha} .
\footnote{ Greek indices $\mu \in \{t,x,y,z\}$
 refer to  boundary directions.}
\end{equation}
Since we are interested in firstly computing $F^{\mu}$, we need to an object that behaves covariantly under the Lorentz transformations
on the boundary. From \eqref{10'} and $P^{M}_{\alpha}=\partial_{\alpha}x^{M}$, it is clear that $ \Pi^{\mu}_{\alpha}\; n^{\alpha}$ has not
our desired property. Instead, $ P^{\mu}_{\alpha}\; n^{\alpha}$ transforms covariantly and we can write:
\begin{equation}\label{12}
F^{\mu}=\;\gamma (\frac{dE}{dt},\frac{d\overrightarrow{p}}{dt})=\;\frac{1}{2\pi\alpha'}\; P^{\mu}_{\alpha}\; n^{\alpha}.
\end{equation} 
Our gauge choice explained above \eqref{9} leads to $n_{\alpha}=(0,-1)$. So the only thing which we need to compute is the value
of $ P_{\mu}^{r}$ on the boundary.

Our strategy is to solve \eqref{10} for three spacial boundary directions $i \in \{x,y,z\}$. 
Since  the Lagrangian of the string includes  $G_{MN}$, $X_{i}'(r,t)$ and $\dot{X}_{i}(r,t)$
\footnote{In this note, prime and dot stand for the derivative with respect to 
$r$ and $t$.}(see \eqref{lagrang}), we can choose one of the spacial directions, i.e. $m$,
and write the Lagrangian in the following simple form:
\begin{equation}\label{14}
\mathcal{L}=\sqrt{A_{m} X_{m}^{'2} +2 B_{m} X_{m}^{'}+C_{m}}\,.
\end{equation} 
Expectedly, $A_{m}$, $B_{m}$ and $C_{m}$ are expressed in terms of $\dot{X}_{m}(r,t)$, $\dot{X}_{\tilde{i}}(r,t)$
and ${X}_{\tilde{i}}^{'} (r,t)$ where $\{\tilde{i}, \tilde{j}, ...\}$ indices refer to the other two spacial directions \eqref{coef}.
This choice of notation extremely simplifies our next computations. Instead of solving EoM for each of $m$ directions, we derive 
 $X^{'}_{m}$ from  the $\Pi^{r}_{m}=\partial \mathcal{L}/\partial x^{'}_{m}$:
\begin{equation}\label{15}
X^{'}_{m}=-\frac{B_{m}}{A_{m}}\pm\frac{\pi_{m}}{A_{m}}\sqrt{\frac{B_{m}^2-A_{m}C_{m}}{\pi_{m}^2-A_{m}}}\,,\;\;\;\;\;\;\pi_{m}=\;-2\pi \alpha'\;\Pi^{r}_{m}\,.
\end{equation}
Then we demand the right hand side to be defined in the whole range of the radial coordinate from the boundary to the horizon. This requirement
 fixes the value of $\Pi^{r}_{m}$ at just one depth in the bulk, namely the world-sheet horizon.  Having the value of $\Pi^{r}_{m}$ at the world-sheet horizon,
 one can integrate equation \eqref{10} to find $\Pi^{r}_{m}$ and thereby $P^{r}_{m}$ on the boundary.
 
 In the next subsection we briefly explain how to compute the drag force in a thermal plasma. Such computation will be essentially the same 
 as computing drag in the fluid dynamics regime at zero order in derivative expansion.
\subsection{Quark in a globally boosted thermal plasma}
\label{plasma}
Consider a thermal state in a $\mathcal{N}=4$ SYM plasma which has been boosted with a global velocity $u^{\mu}$. 
The gravity dual to this state in AdS side may be written as:
\begin{equation}\label{16}\begin{split}
&ds^2 = G_{MN}dx^{M}dx^{N} =\frac{ \mathrm{d}r^2}{r^2 f(br)} +r^2(P_{\mu\nu} -f(br)u_{\mu} u_{\nu}) \mathrm{d} x^{\mu} \mathrm{d}x^{\nu}\,, \\
& f(r)=1-\frac{1}{r^4}\,,         \,\,\,\,\,\,\,\,\,\,   b=\frac{1}{\pi R^2 T}
\end{split}
\end{equation}
where,  $p_{\mu \nu}=u_{\mu}u_{\nu}+\eta_{\mu \nu}$ is the projector tensor orthogonal to  $u^{\mu}$
 and $\eta_{\mu \nu}=(-1,1,1,1)$ is the flat boundary metric  \footnote{$T$ is the Hawking temperature
 of the AdS black brane and  $R$ is the radius of AdS which from now on, we get it equal to one in our
 formulas.}. In a Lorentz frame where plasma  is at rest, $u^{\mu}=(1,0,0,0)$ and so \eqref{14} changes
 to the familiar  form  of an  AdS5 black brane metric \cite{Herzog:2006gh,Gubser:2006bz}. It should be 
 noted that the metric given above is a  four parameter family of solutions for the Einstein equations.
 
 For our next requirements, it is necessary to change the coordinates to the Eddington-Finkelstein coordinates 
 and implement the computations therein\footnote{ In \cite{Abbasi:2012qz}, it has been discussed in detail  the importance of
 choosing this coordinates.}. The metric \eqref{16} in the Eddington-Finkelstein
 coordinates takes the following form:
 \begin{equation}\label{17}
 ds^2 = G_{MN}dx^{M}dx^{N} =-2u_{\mu}dx^{\mu}dr +r^2(P_{\mu\nu} -f(br)u_{\mu} u_{\nu}) \mathrm{d} x^{\mu} \mathrm{d}x^{\nu}.
 \end{equation}
 We are interested in studying a quark in its RF where $\tilde{u}^{\mu}=(1,0,0,0)$. So  $u^{\mu}$
 is the global velocity  of the plasma in the RF of the quark. For the sake of simplicity we also
  assume $u^{\mu}=(u^{0},0,0,u^{z})$.
  In order to solve the EoMs, we should attend to this point that under transforming \eqref{16} to \eqref{17}, all boundary 
  coordinates remain unchanged and only the time coordinate in the bulk transforms.
  Accordingly, we must change the world-sheet time to Eddington-Finkelstein time as well.
 
 \begin{table}\
\centering
\begin{tabular}[h]{|c|c|c|}
\hline
Direction & Components of $F^{\mu}$ & The string profile\\
\hline
\hline
 transverse & $F^{x}=\;F^{y}=\;0$ & $X_{x}(r)=X_{y}(r)=0$  \\
 \hline
  boost & $F^{z}=\;\frac{1}{2\pi \alpha'}\;\frac{u_{z}}{b^2}$& $X_{z}(r)=\;u_{z} b \; \left(\arctan(b r)-\frac{\pi}{2}\right)$\\
\hline
\end{tabular}
\caption{Review of the results for a quark in its RF in a thermal boosted plasma (in the Eddington-Finkelstein coordinates).}
\label{tdpvfd}
\end{table}

 In \eqref{glob} we have discussed in detail how to compute the components of the drag force in a globally boosted thermal plasma.
 We have presented the results in Table 3 \footnote{We have assumed that quark is 
 located at $\overrightarrow{x}=\;0$ on the boundary.}.

%
 %
%
%
 
\section{Drag force in the hydrodynamic regime of $\mathcal{N}=4$ SYM plasma}
\label{fluid}
After determining the most general form of the covariant drag force in section\eqref{most}, our next goal is to determine
the unknown coefficient functions in \eqref{6} in the $\mathcal{N}=4$ SYM field theory.
On the other hand, Fluid/Gravity duality provides the dual gravity of a general fluid dynamical flow in this field theory.
So in this section we first briefly review the the Fluid/Gravity duality and then explain in detail how to implement perturbative 
computations on a classical string in gravity side.

\subsection{Review of the Fluid/Gravity duality}
\label{fluid/gravity}  
 Fluid dynamics studies long-wavelength peturbations in a thermal system.
This feature can be captured in  local derivative expansion of physical quantities , i.e. energy density, pressure, etc..
 The order of every term  in this expansion is determined  by  number of its derivatives. Roughly speaking, the ratio of 
 every term in a definite order  to the terms in one lower order,  is of the order of $1/L T$,  with $L$ the scale over
  which the thermodynamic variables vary significantly. The system will be in local  equilibrium state if
 \begin{equation}\label{23}
 \frac{1}{L T}\ll1.
 \end{equation}
In this limit the scale of variations is so long that in each of  fluid patches\footnote{A region with the size of effective $l_{mfp}\sim 1/T$.} 
thermodynamic is dominant. So it makes sense  to promote the  equilibrium degrees of freedom\footnote{In 4-dim there are four of them, $T$ and three of $u^{\mu}$,
($u^{\mu}u_{\mu}=-1$).} to become local functions of space and time. Local conservation of energy-momentum tensor and
 thermodynamic equation of state will  completely describe the evolution of the system in the limit \eqref{23}.
 This is ideal fluid description or zero order fluid dynamics.

 The greater the above ratio is, the more deviation from local equilibrium  would  arise. So it will be needed to at least take the one derivative terms  into
  account in the expansion. In more quickly varying flows where $1/LT$ becomes greater, one also might have to go beyond the first
  order  and correct the expansion by adding terms with  more than one derivative.

 In \cite{Bhattacharyya:2008jc}  it has been shown that  there is a
one to one map between fluid dynamical flows of a $\mathcal{N}=4$ SYM field theory on the boundary 
and long-wavelength perturbations of an AdS5 black brane in the bulk. 
 This duality, i.e. the Fluid/Gravity duality, is able to construct the five dimensional gravitational background 
dual to every given four dimensional boundary flow, perturbatively order by order in a boundary derivative expansion.

\subsubsection{Dual gravity of a boundary flow}
\label{aspects}
Fluid/Gravity duality has been  originally constructed  for a general fluid dynamical
 flow on the boundary without using its profile.
 The idea is to restrict the computations to be implemented within just one boundary patch. Fixing the value of the temperature and velocity at one point, 
 one can find the fluid profile over the whole patch points via Taylor expansion. On the other hand using Eddington-Finkelstein coordinates
 allows to  extend this boundary patch into the bulk through a tube-wise region and so all computations in gravity side reduces to
 solving Einstein equations  in this tube perturbatively, for full range of the radial coordinate. The final step is
 to covariantize the obtained metric in boundary directions.    

Requiring the slowly varying condition on $u^{\mu}(x)$ and $T(x)$, one may expect every tube-wise region  to become a local
black brane solution \eqref{17}, at zero orde:
\begin{equation}\label{24}
ds_{(0)}^2 = G^{(0)}_{MN}dx^{M}dx^{N} =-2u_{\mu}(x^{\alpha})drdx^{\mu}+ \frac{r^2}{R^2}(P_{\mu\nu}(x^{\alpha}) - f(b(x^{\alpha}) r)u_{\mu}(x^{\alpha})u_{\nu}(x^{\alpha}))dx^{\mu}dx^{\nu}
\end{equation} 
where $P_{\mu\nu}(x^{\alpha})=u_{\mu}(x^{\alpha})u_{\nu}(x^{\alpha})+\eta_{\mu\nu}$.

Going to next orders in perturbation, $u^{\mu}(x^{\alpha})$ and $T(x^{\alpha})$ will not be constant over
the patch points though. In any order of perturbation, we have to add some appropriate derivative terms to
 the metric to make it a tube-wise solution  in the bulk to the given order.  Note that added  terms are generically
 constructed by the derivatives of  $u^{\mu}$ and $T$.
The first order corrections to the metric, $G_{MN}^{(1)}$ elements, are given by:
\begin{equation}  \label{26}
\mathrm{d}s^2_{(1)}= r^2 \,b\, F(b\, r)\, \sigma_{\mu\nu}  \, dx^{\mu} dx^{\nu} 
+{2\over 3} \, r \, u_{\mu}u_{\nu} \,\partial_{\lambda} u^{\lambda} \, dx^{\mu}dx^{\nu} -  r\, u^{\lambda}\partial_{\lambda}\left(u_\nu u_{\mu}\right)\, dx^{\mu} dx^{\nu}   
\end{equation}
where
\begin{equation}\label{27}
F(r) =\;{1\over 4}\, \left[\ln\left(\frac{(1+r)^2(1+r^2)}{r^4}\right) - 2\,\arctan(r) +\pi\right] 
\end{equation}
and 
\begin{equation}\label{28}
\sigma^{\mu\nu}= P^{\mu \alpha} P^{\nu \beta} \, 
\, \partial_{(\alpha} u_{\beta)}
-\frac{1}{3} \, P^{\mu \nu} \, \partial_\alpha u^\alpha. 
\end{equation}
In next subsection we argue how to use \eqref{24} and \eqref{26} to implement perturbative 
computations in a patch.
  \subsubsection{Nature of derivative expansion}
  \label{nature}
Let us restate that in hydrodynamic expansion, the order of every term is determined by number of its derivatives.
In order to explicitly observe  the order of different terms, it would be useful to regard $u$ and $b$ as functions of $\epsilon\; x^{\alpha}$
where $\epsilon$ is a parameter that will be finally set to be one. Since every derivative of $u$ and $b$ produces a power of $\epsilon$,
 powers of $\epsilon$ count the order of terms in the expansion.
  
  According to our discussion in previous subsection, we have to perform our computations  within one patch.
  The reason for this is that we are studying the drag force in a general flow with unknown dynamical profile. 
  Practically we must express every dynamical term in a Taylor expansion  around an arbitrary point in the patch. Let us clarify it
  by an example. Suppose we want to expand a  metric element, $H$, in derivative expansion up to first order.
  In contrast to the AdS black brane case, $H$ is not constant here and  generically is a function of boundary coordinates through fluid variables, 
  $\lambda_{p}(t,\overrightarrow{x})=\{u^{\mu}(x^{\alpha}),\;b(x^{\alpha})\}$\footnote{We have labelled different variables with $p$.},
  and also of radial coordinate, $r$. So $H$ may be expanded in a patch as following\footnote{Notice that  we have expanded $H$ around the point $(t,\overrightarrow{x})=(0,0)$.}:
  \begin{equation*}\label{29}
  \begin{split}
  H(\lambda_{p}(t,\overrightarrow{x}),r)&= H^{(0)}(\lambda_{p}(0,0),r)\\
  &+\underbrace{\epsilon\;(\overrightarrow{x}\partial_{\overrightarrow{x}}+t\partial_{t})\lambda_{p}(0,0)\;\;\frac{\partial{H^{(0)}}}{\partial \lambda_{p}}(\lambda_{p}(0,0),r)}_{Taylor}+\;\underbrace{\epsilon \frac{}{}H^{(1)}(\lambda_{p}(0,0),r)}_{Hydrodynamic}+O(\epsilon ^{2}). 
  \end{split} 
  \end{equation*}
  In the  first line we have demonstrated the zero order part of $H$
that is exactly the same as $H$ in an AdS black brane metric with  $\lambda_{p}=\;\lambda_{p}(0,0)$. This may simply 
recall that zero order fluid dynamics in a patch is nothing except thermodynamics.
The first order corrections of $H$ have  been written  in the second line; while the first term comes
from the Taylor expanding, the second one is a non trivial pure hydrodynamic correction. Such non trivial correction is a combination of
 $\partial \lambda_{p}$ terms  which can be specified via the Fluid/Gravity duality. 

We can resume Taylor expansion and write $H$ in the following compact form:  
\begin{equation}\label{29'}
 H(\lambda_{p}(t,\overrightarrow{x}),r)=\;H^{(0)}(\lambda_{p}(t,\overrightarrow{x}),r)+\;\epsilon\;H^{(1)}(\lambda_{p}(0,0),r)+\;O(\epsilon ^{2})
\end{equation}
  The really important point with this statement is that the first term does not have to be taken into account as a fully local term. It must be only regarded 
  as a local function of $x^{\alpha}=(t,\overrightarrow{x})$ up to first order in Taylor expansion. So from now on, we make the following convention:
  We mean by every local function, the summation of zero and first order terms in its Taylor expansion.

\subsection{Quark in the hydrodynamic regime of $\mathcal{N}=4$ SYM plasma}
\label{method}
Let us recall that as far as the string profile in the bulk  does not extend  beyond a tube,
one can proceed to perturbatively solve the string EoMs.  In \cite{Abbasi:2012qz} we showed that using Eddington-Finkelstein coordinates,
the string always lies in one tube and perturbative computations is meaningful.

\subsubsection{Expanding the metric to first order}
\label{metric}
Now we exploit what we have explained in previous subsection to rewrite the first order corrected metric.
Suppose at $t=0$ at  the quark position, $\overrightarrow{x}=0$, in its RF,  fluid velocity is given by:
\begin{equation}\label{30}
u^{\mu}(0,0)=\;(u^{0},0,0,u^{z});
\end{equation}
so the corrected metric in the neighbourhood of the quark can be written as:
\begin{equation}\begin{split}\label{31}
ds^2=&-2u_{0}\;dr dt-\;2u_{z}\;dr dz
+\;\frac{u_{0}^2}{r^2 b^4}\;dt^2+\;\frac{u_{z}^2}{r^2 b^4}\;dz^2+\;\frac{u_{0}u_{z}}{r^2 b^4}\;dt dz+r^2\eta_{\mu \nu} dx^{\mu}dx^{\nu}\\
&+\;G_{M N}^{(Taylor)}\; dx^{M} dx^{N}+\;G_{M N}^{(1)} \;dx^{M} dx^{N}
\end{split}
\end{equation}
As a useful point, in next subsection we will show in detail that it will be possible to implement the computations without dealing with "Taylor" corrections.
Before proceeding we discuss about $G_{MN}^{(1)}$ elements now. These are pure hydrodynamic corrections that can be simply computed via \eqref{26}
with applying the  useful expressions given below:
 \begin{equation}\label{32}
  \begin{split}
  &\sigma_{tz}=\;\frac{u_{z}}{u_{0}}\;\partial_{z}u_{z}+u_{z}\;\partial_{t}\; ln \;b\\
  &\sigma_{tt}=\;\frac{u_{z}^2}{u_{0}^2}\;\partial_{z}u_{z}-\;\frac{u_{z}^2}{u_{0}}\;\partial_{t}\;ln \;b.
  \end{split}
  \end{equation}
  There is a delicate point with the expressions given above. In the Fluid/Gravity  duality, derivative corrections at first order have
   been introduced in terms of  velocity derivatives. The temperature derivatives have been eliminated in favour of them
    via using fluid dynamics EoMs. However as it is clear in Table 2 we may deal with temperature derivative terms throughout our
    computations, i.e. $s_{2}$.
    Hence we must reuse EoMs to restore such terms in all possible expressions. Using 
  \begin{equation}\label{321}
\partial_{\mu}T^{\mu t}_{(0)}=0,\;\;\;\;
\partial_{\mu}T^{\mu z}_{(0)}=0
  \end{equation}
one can simply show:
\footnote{Notice that $\sigma_{tz}$ and $\sigma_{tt}$ are not involved with the derivative terms in the other two equations
    $\partial_{\mu}T^{\mu x}_{(0)}=0$ and $\partial_{\mu}T^{\mu y}_{(0)}=0$.}
 \begin{equation}\label{322}
 \begin{split}
  &\partial_{z}b=\;\frac{b}{u_{0}^2}\;u^{\alpha}\partial_{\alpha}u^{z}+\;\frac{u_{z}}{u_{0}}\;\partial_{t}b\\
  &\partial_{\mu}u^{\mu}=\;\frac{3}{u_{0}}\;\left(\frac{u_{z}}{u_{0}}\;u^{\alpha}\partial_{\alpha}u_{z}-\;\partial_{t}\;ln \;b \right).
      \end{split}
  \end{equation}
This is why we see only $\partial_{t} b$ and $\partial_{\alpha} u_{z}$ in \eqref{32}. Respecting all considerations mentioned
above, we will compute the drag force in the background with metric \eqref{31}.

Before start trying to compute the drag force, we will briefly discus about  structure of our
perturabtive method in next subsection.

\subsubsection{Structure of the perturbative method}
\label{structure}
In order to compute the drag force in the hydrodynamic regime we  should first expand $A_{m}$, $B_{m}$
and $C_{m}$ in \eqref{15} up to first order in derivative.
This expanding includes both metric elements and  components of the string profile, see \eqref{coef}. 
In  \eqref{29} we explained how to expand metric elements in a patch. However the situation is a little 
bit different in the case of the string profile.
Let us clarify this point now. We may naturally expect:
\begin{equation}\label{33}
X_{i}(r,t)=\;X^{(0)}_{i}(r,t)+\;\epsilon\;X^{(1)}_{i}(r,t)+\;O(\epsilon^2)
\end{equation}
where the process of generating zeroth order solution of the string is the same as producing zero order metric \eqref{24}.
We must get the solution of the string in a thermal system and replace thermodynamic variables with local hydrodynamic functions
in it, i.e. :
\begin{equation}
\begin{split}\label{34}
&X_{x}^{(0)}(r,\;t)=\;0\\
&X_{y}^{(0)}(r,\;t)=\;0\\
&X_{z}^{(0)}(r,\;t)=\;u_{z}(t,X_{i})\; b(t,X_{i}) \;\left[\arctan(b(t,X_{i}) r)-\frac{\pi}{2} \right].
\end{split}
\end{equation}
The very important point with this expression is the presence of $X_{z}(r,t)$ not only in left hand side,
but also in the argument of the hydrodynamic variables in right hand side! Notice that $X_{i}(r,t)$ in right hand side is
 a function of $u_{z}(t,X_{i})$ and $b(t,X_{i})$; $X_{i}$ itself; in the argument of these variables is also function of 
 $u_{z}(t,X_{i})$ and $b(t,X_{i})$ and so on. Hence this equation behaves as a recursive formula
for $X_{z}(r,t)$. However demanding $X_{z}(r,t)$ to be computed at zero order simply truncates this infinite recurrence.

According to discussion below \eqref{29'}, the locality of \eqref{34} is valid up to the first order in Taylor expansion around
every arbitrary point. It means that whenever \eqref{34} is considered in a patch, it will also include a first order part.
So  up to first order in derivative expansion we may write: 
\begin{equation}\label{35}
X_{z}(r,t)=\underbrace{u_{z}(0,0)\;  b(0,0) \left[\arctan(b(0,0) r)-\frac{\pi}{2}\right] }_{X_{z}^{(0)}(r)}+\;\epsilon\; X_{z}^{(Taylor)}(r,t)+\;\epsilon\;X_{z}^{(1)}(r,t)+O(\epsilon^2)
\end{equation}
In both one derivative corrections above, the hydrodynamic variables and their derivative can be computed at any arbitrary point, for instance at $(0,0)$.
\textbf{Our strategy is to perturbatively compute  $X_{z}^{(1)}(r,t)$ via correcting the location of point at wich $B^2-A C$ vanishes} \footnote{At zero order, this point coinsides with the position of world-sheet horizon.}.

As we described in section\eqref{plasma}, the  expression $B_{m}^2-A_{m}C_{m}$ has a simple root, $r^{*}$, in the bulk of AdS black brane.
Our main idea in hydrodynamic regime is that $r^{*}$ is  perturbatively corrected order by order in derivative expansion.
In what follows we try to clarify some important points with perturbatively solving $E_{m}=\;B_{m}^2-A_{m}C_{m}=0$.

In the language of section\eqref{nature}, we are interested in finding $r^{*}$, the  root of $E_{m}$. So
we must compute $r^{*}(t)$ in the following equation up to first order:
\begin{equation}\label{36}
E _{m}\left(\lambda_{p}(t,X_{i}^{*}(r^{*}(t),t))\;,\frac{}{}r^{*}(t) \right)=\;0.
\end{equation}
 It should be noted that the time dependency of $r^{*}$ is originated from its dependence on fluid variables, i.e. $\lambda_{p}(t,\overrightarrow{x})$.
 Carefully rewriting:
 \begin{equation}\label{37}
 r^{*}(t):=\;r^{*}\left(\lambda_{p}(t,X_{i}^{*}(r^{*}(t),\frac{}{}t))\right)
 \end{equation}
where  $r^{*}(t)$ is clearly present in  both sides. As a result, this equation only expresses $r^{*}(t)$ implicitly.
However  up to first order in derivative expansion we can derive it as:
\begin{equation*}\label{38}
r^{*}(t)=\;r^{*}_{(0)}(\lambda_{p}(0,0))+\epsilon\;(X_{i}^{*}\partial_{X_{i}^{*}}+\;t\partial_{t})\lambda_{p}(0,0)\;
\frac{dr^{*}_{(0)}}{d\lambda_{p}}(\lambda_{p}(0,0))+\;\epsilon\;r^{*}_{(1)}(\lambda_{p}(0,0))+\; O(\epsilon^2).
\end{equation*} 
In this equation $\;r^{*}_{(0)}(\lambda_{p}(0,0)):=r^{*}_{(0)}$ is nothing except the $r^{*}$  introduced 
below \eqref{18} whose $\lambda_{p}$terms have been replaced with $\lambda_{p}(0,0)$.
Now we  plague $r^{*}(t)$ in \eqref{36} and expand $E_{z}$ up to first order:
\begin{equation}\label{40}
\begin{split}
E_{z}=&\;E_{z}^{(0)}(\lambda_{p}(0,0),\;r^{*}_{(0)}(0))\;\textbf{+}\;\epsilon\;(X_{i}^{*}\partial_{X_{i}^{*}}+\;t\partial_{t})\lambda_{p}(0,0)\;\;\;\frac{\partial E_{z}^{(0)}}{\partial\lambda_{p}}(\lambda_{p}(0,0),r^{*}_{(0)}(0))\\
&\textbf{+}\;\epsilon\;(X_{i}^{*}\partial_{X_{i}^{*}}+\;t\partial_{t})\lambda_{p}(0,0)\;\;\;\frac{dr^{*}_{(0)}}{d\lambda_{p}}(\lambda_{p}(0,0))\;\;\;\frac{\partial E_{z}^{(0)}}{\partial r^{*}}(\lambda_{p}(0,0),r^{*}_{(0)}(0))\\
&\textbf{+}\;\epsilon\;r^{*}_{(1)}(0)\;\;\; \frac{\partial E_{z}^{(0)}}{\partial r^{*}}(\lambda_{p}(0,0))\;\textbf{+}\;\epsilon\;E_{z}^{(1)}(\lambda_{p}(0,0),r^{*}_{(0)}(0))\;\textbf{+}\;O(\epsilon^2).
\end{split}
\end{equation} 
As it is clear, the second term in the first line and the term in the second line are first  order Taylor terms. Interestingly, they cancel out each other
\footnote{We have used this fact that from a relation like $F(x,y)=0$ one may deduce  $y'_{x}=\;-\;\partial_{x}F(x,y)/\partial_{y}F(x,y)$.}.  $E_{(0)}$  as the only zero order
term in $E_{z}$ gives $r^{*}_{(0)}=\;r^{*}_{(0)}(\lambda_{p}(0,0))=$ constant. The remaining two terms in the third line determine $r^{*}_{(1)}$.

In summary, we need only to work out the value of $B_{m}^2-A_{m}C_{m}$ and its pure hydrodynamic correction at the position of the quark.
Setting $E_{z}$ to zero, the former one gives  $r^{*}_{(0)}$ and  $r^{*}_{(1)}$ can be computed via hydrodynamic correction part.
  
  \subsubsection{Drag force in the boost direction}
  \label{boost}
  Let us denote that by the boost direction here, we mean the direction of the fluid velocity at the quark position in its RF.
Having used  \eqref{10'} and \eqref{12}, it is clear that 
  \begin{equation}
  F^{z}=\;F_{z}=\;\lim_{r\rightarrow \infty}P^{r}_{z}=\;\lim_{r \rightarrow \infty} \left(\frac{1}{\sqrt{-g}}\;\Pi^{r}_{z} \right).
  \end{equation}
To compute the covariant drag in the boost direction we only need to know the value of   $\Pi^{r}_{z}$ on the boundary.

  Following our explanations in last subsection,  we try to specify zero and first order parts of $A_{z}$,  $B_{z}$
  and $C_{z}$ with $\lambda_{p}$ terms computed at the  quark position. 
Instead of listing these coefficients here we directly proceed to compute $B_{z}^2-A_{z}C_{z}$ with $u$ and $b$ computed
at the position of the quark (see \eqref{first order coef} for more details.). Setting \eqref{40} 
to zero, $r^{*}_{(0)}$ and $r^{*}_{(1)}$ turn out to be as the following:
\begin{equation}\label{41}
\begin{split}
&r^{*}_{(0)}=\;\frac{\sqrt{u^{0}}}{b}\,,\\
& r^{*}_{(1)}=\;\frac{1}{4\;r^{*3}_{(0)}}\;\left[\frac{2 u_{0} u_{z}}{b^4} \dot{X}_{z}^{(0)}(r^{*}_{(0)})+\;\frac{u^{0}}{b^2}G_{tt}^{(1)} (r^{*}_{(0)})\right]\,.
\end{split}
\end{equation}
   Now we demand \eqref{15} to be well-defined for the whole range of the radial coordinate. It simply  results in fixing the value of $\pi_{z}$ at
    $r^{*}_{(0)+(1)}(t)$, up to first order. Considering the true direction for the energy and momentum flow on the 
   world-sheet \cite{Abbasi:2012qz}, we may write:
    \begin{equation}\label{42}
    \pi_{z}\left(r^{*}(t),\frac{}{}\lambda_{p}(t,X^{*}_{i}) \right)=\;-\sqrt{A_{z}^{(0)}(r^{*}_{(0)})}\;-\;\epsilon\;\frac{A^{(0)}_{z(Taylor)}(r^{*}_{(0)})+\;r^{*}_{(1)}\;A_{z}^{(0)'}(r^{*}_{(0)})\;+\;A_{z}^{(1)}(r^{*}_{(0)})}{2\sqrt{A_{z}^{(0)}(r^{*}_{(0)})}}
    \end{equation}
 where
 \begin{equation}\label{43}
 A^{(0)}_{z(Taylor)}(r^{*}_{(0)})=\;4 r^{*3}_{(0)}\;\;(X^{*}_{i}\partial_{X_{i}}+t\partial_{t})\lambda_{p}\;\;\frac{dr^{*}_{(0)}}{d\lambda_{p}}\;+\;\frac{4}{b^{5}}\;(X^{*}_{i}\partial_{X_{i}}+t\partial_{t})b
 \end{equation}
  is the only time dependent term in \eqref{42}. According to \eqref{37}, $ \pi(r^{*}(t))$ is a function of the fluid 
  variables computed at the point $(t, X^{*}_{i})$ which is different the quark position. However the drag force is a function of 
  fluid variables computed at the quark position.  To find $\pi_{z}$ at the desired point we should integrate \eqref{10} for $M=z$ from
  $r^*_{(0)+(1)}$  to the boundary:
  \begin{equation}\label{44}
   \Pi_{z}^{r}\left(r,\frac{}{}\lambda_{p}(t,X_{i}) \right)=\; \Pi_{z}^{r}\left(r^{*}(t),\frac{}{}\lambda_{p}(t,X^{*}_{i}) \right)+\;
   \int_{r^{*}(t)}^{r}(f_{z}-\; \partial_{t}\Pi^{t}_{z})\;d \mathbf{r}.
  \end{equation} 
  There are two important points regarding this equation. Firstly, the integral term in right hand side appears just at first order.
 The reason for this is that both terms in the  integrand are boundary derivative terms. So  we should  take into account only 
 the zero  order part of $r^{*}$ in the lower limit of the integral. Secondly,  computing $ \Pi_{z}^{r}$  at the the quark position,
 $X_{i}=0$ at $t=0$,  removes explicit time dependency from \eqref{44} (see  \eqref{43}). 
  
  After rather tedious simplifications one reaches the following formula:
  \begin{equation}\label{45}
  \begin{split}
  F^{z}(\lambda_{p}(0,0))&=\;F^{z}_{(0)}+\;\epsilon\;F^{z}_{(1)}+\;O(\epsilon^2)\\
  &=\;-\frac{1}{2\pi \alpha'}\;\left(-\frac{u_{z}}{b^2}+\;\mathcal{A} \;\partial_{t}u_{z} +\mathcal{B}\;  \partial_{z} u_{z}+\mathcal{C}\; \frac{\partial_{t} b}{b} \right)
  \end{split}
  \end{equation}
 with $u_{0}$, $u_{z}$ and $b$ the fluid variables at $(t,\overrightarrow{x_{i}})=\;(0,0)$ and 
   \begin{equation}\label{46}
  \begin{split}
 \mathcal{A}&=\frac{1}{b}\frac{\sqrt{u^{0}}}{u^{0}}\,,\\
u_{0}\, \mathcal{B}=\, -b\,\mathcal{C}&=\;\frac{1}{b}\frac{\sqrt{u^{0}}}{u^{0}}\;u_{z}+\;\mathcal{U} \;\frac{1}{b} \frac{(1+u_{0}^2)}{u^{0}}u_{z}\;-\;\frac{F(\sqrt{u^0})}{b}\frac{(1+u_{0}^2)}{u^{0}}\;u_{z}\,,\\
 \mathcal{U}&=\;\;\frac{\pi}{2}-\arctan(\sqrt{u^{0}})\,.
\end{split}
\end{equation}
One might  worry about the absence of the other derivative terms in \eqref{45}, like $\partial_{x}u_{z}$.
The point is that we have used the fluid dynamics equation of motion to eliminate such terms.  In our computations, the only
source of  these terms is $\partial_{\mu}u^{\mu}$ appearing in metric corrections.
It is pretty simple to replace such term with the above-mentioned derivatives in \eqref{45} by use of \eqref{322}.

In \eqref{middle}, we have listed all intermediate steps related to the computations of this subsection.
  
  \subsubsection{Drag force in the transverse directions}
  \label{transverse}
  Notice that by  the transverse directions here, we mean the directions orthogonal to the fluid velocity at the quark position in its RF.
Similar to the boost direction, the only thing that we need to compute related to these directions is the value of   $\Pi^{r}_{x,y}$ on the boundary.
 We have to firstly know the zero and first order parts of $A_{T}$, $B_{T}$ and $C_{T}$
with fluid variables computed at $(t,\overrightarrow{x_{i}})=\;(0,0)$ (T stands for the transverse directions; i.e. x,y.). The crucial point that
extremely simplifies the expansion of these coefficients is vanishing of the string profile in the transverse directions at zero order (See
\eqref{middle T} for more details.). Consequently it can be shown that the drag force in transverse directions can be computed by use of just 
zero order parts of these coefficients listed below:
\begin{equation}\label{48}
\begin{split}
&A^{(0)}_{T}(\lambda_{p}(0,0))=\;r^4 f(b r)-\;\frac{u_{z}^2}{b^4}\,,\\
&B^{(0)}_{T}(\lambda_{p}(0,0))=\;0\,,\\
&C^{(0)}_{T}(\lambda_{p}(0,0))=\;1\;.
\end{split}
\end{equation}
Like the $z$ direction case, we assume  $B^2_{T}-A_{T}C_{T}$  vanishes at $r_{T}^{*}=\;r^{*}_{T(0)}+\;\epsilon \;r^{*}_{T(1)}$.
It is simple to find $r^{*}_{T(0)}$ and $r^{*}_{T(1)}$ via expanding $B^2_{T}-A_{T}C_{T}$ in the form of \eqref{40}.
As it was explained below \eqref{40}, Taylor contributions would precisely cancel. Solving $E_{T}=\;0$ at zero and first order separately, one 
simply reaches to\footnote{Interestingly, by use of \eqref{A15}, it is pretty simple to show that
	\begin{equation}\label{491}
	r^{*}_{T(1)}=\;r^{*}_{(1)}.
	\end{equation}
	this equality can be regarded as a very good check of our computations.}:
  \begin{equation}\label{49}
  \begin{split}
 &r^{*}_{T(0)}=\;\frac{\sqrt{u^{0}}}{b}\,,\\
&r^{*}_{T(1)}=\;-\frac{A_{T}^{(1)}}{4\;r^{*3}_{T(0)}}\,.
  \end{split}
  \end{equation}

Now, we compute  
 $\pi_{T}\left(\tilde{r_{T}}^{*}(t),\frac{}{}\lambda_{p}(t,X^{*}_{i}), \right)$ somewhat tricky. Consider the following relation for the transverse momentm current on the world-sheet:
  \begin{equation}
  	\pi_{\perp}=\,\frac{A_{\perp} X'_{\perp}+\,B_{\perp}}{\sqrt{g}}
  \end{equation}
  Since  $B_{\perp}^{(0)}=0 $ and  $X_{\perp}^{(0)}=0 $, the numerator is at least of first. Therefore the denominator has to be computed at zero order i.e. $g^{(0)}=1$.
  So to first order in derivative expansion we may write:
    \begin{equation}\label{52.5}
    	\begin{split}
    		\pi_{\perp}(r^{*}_{\perp(0)})&=\,B_{\perp}^{(1)}\\
    		&=\left(u_{0}u_{z}^{2}\frac{F(\sqrt{u^{0}})}{b}+\,u_{0}^{2}\frac{\sqrt{u^{0}}}{b}+\,u_{0}\frac{\mathcal{U}}{b}\right)\,\partial_{t}u_{\perp}\\
    		&+\left(-u_{0}^{2}u_{z}\frac{F(\sqrt{u^{0}})}{b}-\,u_{0}\frac{\sqrt{u^{0}}}{b}+\,u_{z} \frac{\mathcal{U}}{b} \right) \partial_{z}u_{\perp}\\
    		& -\,u_{0}\frac{F(\sqrt{u^{0}})}{b} \,\partial_{\perp}u_{0}
    	\end{split}
    \end{equation}
    Computing the right hand side of the equation of motion:
    
    \begin{equation}\label{53.5}
    	\int_{r^{*}_{\perp (0)}}^{\infty}(f_{\perp}-\; \partial_{t}\Pi^{t}_{\perp})\;d \mathbf{r}\;=\,-\frac{1}{2 \pi \alpha'}\frac{u_{0}}{b}\,\mathcal{U}\,\partial_{t}u_{\perp},
    \end{equation}
    one can finally give the transverse drag force as it follows:
    \begin{equation}\label{54.5}
    	\begin{split}
    		F^{\perp}(\lambda_{p}(0,\vec{0}))&=\;F^{\perp}_{(0)}+\;\epsilon\;F^{\perp}_{(1)}+\;O(\epsilon^2)\\
    		&=\;-\frac{1}{2\pi \alpha'}\;\left(\mathcal{D} \;\partial_{t}u_{\perp} +\mathcal{E}\;  \partial_{z} u_{\perp}+\mathcal{F}\; \partial_{\perp} u_{0} \right)
    	\end{split}
    \end{equation}
where:
    \begin{equation}\label{46}
    	\begin{split}
    		&\mathcal{D}=u_{0}u_{z}^{2}\frac{F(\sqrt{u^{0}})}{b}+\,u_{0}^{2}\frac{\sqrt{u^{0}}}{b}\,,\\
    		&\mathcal{E}=-u_{0}^{2}u_{z}\frac{F(\sqrt{u^{0}})}{b}-\,u_{0}\frac{\sqrt{u^{0}}}{b}+\,u_{z} \frac{\mathcal{U}}{b} \,,\\
    		&\mathcal{F}=\;-u_{0}\frac{F(\sqrt{u^{0}})}{b} \,,\\
    		&\mathcal{U}=\;\;\frac{\pi}{2}-\arctan(\sqrt{u^{0}})\,.
    	\end{split}
    \end{equation}
 
 Let us summarise. In this section we studied a quark moving through  a general flow of $\mathcal{N}=4$ SYM field theory.
 Our main achievement was computing the components of the drag force  exerted on quark, perturbatively, up to first order
 in derivative expansion. Our  results were specially obtained in the quark's RF at $t=0$   once the fluid velocity at the  
 quark position  directed along $z$ direction. 
  In order to express the drag force in a general frame at next times, in next section we covariantize the results obtained in this section.

 \section{ Covariant drag force in  $\mathcal{N}=4$ SYM field theory}
 \label{matching}
In section\eqref{most}, we presented the most general form of first order corrected  covariant drag
 force in a general fluid flow with six unknown coefficient functions. Now,  we fix
  those coefficient functions in the $\mathcal{N}=4$ SYM field theory, by use of the results  obtained
 in previous section.
  
  \subsection{Specifying the coefficient functions in $F^{\mu}$}
  \label{specify}
  In order to compare the general covariant formula with the results of section\eqref{fluid}, we have to first rewrite
   the components  of general $F^{\mu}$  under the same conditions where we obtained   \eqref{45} and
  \eqref{54.5}. Therefore we require components of $F^{\mu}$ in the quark RF  at $(t,\overrightarrow{x})=(0,0)$, 
  assuming $u^{\mu}_{(0,0)}=(u^{0},0,0,u^{z})$ and $\tilde{u}^{\mu}=(1,0,0,0)$. Denoting all these consideration
  and information listed in Table 4, we write  spacial components of general $F^{\mu}$ at the quark position at $t=0$\:

\begin{table}
\centering
\begin{tabular}[h]{|c|c|c|}
\hline
\hline
SO(3) Type & In dependent data &Evaluated in RF \\
\hline
\hline
Scalars & $s_{1}= \tilde{u}^{\mu} \tilde{u}^{\nu} \partial_{\mu} u_{\nu}$ &$u_{0}^{-1}\;u_{z}\partial_{t}u_{z}$  \\
              & $s_{2}=\tilde{u}^{\mu}\partial_{\mu}b$ &$\partial_{t}b$ \\
              & $s_{3}=u^{\mu} \tilde{u}^{\nu} \partial_{\mu} u_{\nu}$ & $u_{0}^{-1}\;u_{z}(-u_{0}\partial_{t}u_{z}+u_{z}\partial_{z}u_{z})$\\
          \hline
Vectors & $v_{1}^{\mu}=\tilde{p}^{\mu \alpha} \tilde{u}^{\beta} \partial_{\alpha} u_{\beta }$& $\tilde{p}^{\mu \alpha} \partial_{\alpha} u_{0 }$ \\
               & $v_{2}^{\mu}=\tilde{p}^{\mu \alpha} \tilde{u}^{\beta} \partial_{\beta} u_{\alpha }$& $\tilde{p}^{\mu \alpha}  \partial_{t} u_{\alpha }$ \\
               &$v_{3}^{\mu}=\tilde{p}^{\mu \alpha} u^{\beta} \partial_{\beta} u_{\alpha }$& $\tilde{p}^{\mu \alpha}(u_{z}\partial_{z}u_{\alpha}-u_{0}\partial_{t}u_{\alpha})$\\
              \hline
              \hline
              \end{tabular}
              \caption{First order derivative data evaluated in RF.}
\end{table}

\begin{subequations}
\begin{eqnarray}\label{541}
  &&  F^{x}(\lambda_{p}(0,0))=\;\sigma_{1}\;\partial_{x}u_{0}+\;\sigma_{2}\;\partial_{t}u_{x}+\;\sigma_{3}\;\partial_{z}u_{x}\,,  \\
\label{542}  
  &&  F^{z}(\lambda_{p}(0,0))=\;\frac{1}{2\pi \alpha'}\left(1+\frac{}{}\kappa_{1}\;\partial_{t} u_{z}+\kappa_{2}\;\frac{\partial_{t}b}{b}+\kappa_{3}\;\partial_{z}u_{z}\right)\;\frac{u_{z}}{b^2}
  \end{eqnarray}
\end{subequations}

with  
\begin{subequations}
	\label{55}
	\begin{align}
		&\frac{u_{z}}{b^2}\, \kappa_{1}=-\alpha_{3}\;u_{z}+\alpha_{1}\;\frac{u_{z}}{u_{0}}, &
		& \sigma_{1}=\tilde{\beta}_{1}, \\
		&\frac{u_{z}}{b^2} \,\kappa_{3}=\alpha_{3}\;\frac{u_{z}^2}{u_{0}}, &
		& \sigma_{2}=\tilde{\beta}_{2}-u_{0}\;\tilde{\beta}_{3}, \\
		\label{E:sigmaDef}
		&\frac{u_{z}}{b^2} \,\kappa_{2}=\alpha_{2}, &   
		& \sigma_{3}=\tilde{\beta}_{3}\; u_{z} \,.
	\end{align}
\end{subequations}
where fot sake of simplicity, we have used  $\,\beta_{i}=\,-2\pi \alpha'\tilde{\beta_{i} }$.
 
Comparing $F^{\perp}$ obtained in (\ref{54.5})
 and the one expressed through (\ref{541})  we simply find:
     \begin{equation}\label{66.5}    
     \boxed{ 
          \begin{split}
          	 &\tilde{\beta}_{1}=s^2\,\frac{F(s)}{b}\\
          	&\tilde{\beta}_{2}=s^2\,\frac{F(s)-G(s)}{b} \\
          	&\tilde{\beta}_{3}=\,\frac{G(s)-\,s^4\,F(s)+\,s^3}{b}      \end{split}}
          \end{equation}
 with $s=\sqrt{-S_{1}}=\sqrt{-\tilde{u}.u(x)}$ and:
 \begin{equation}
\begin{split}
G(s)&=\;\;\frac{\pi}{2}-\arctan(s)\\
F(s) &=\;{1\over 4}\, \left[\ln\left(\frac{(1+s)^2(1+s^2)}{s^4}\right) - 2\,\arctan(s) +\pi\right]. 
\end{split}
 \end{equation}
 In order to find $\alpha_{i}$ function, we should compare $F^{z}$ presented in \eqref{45} with \eqref{542}.
 The argument in the last paragraph of \eqref{boost} assures us that the three different derivative terms appeared in  \eqref{45}
 are fully independent. Correspondingly, the same derivative terms have been appeared in \eqref{542}. So the comparison between these equations
 results in three equations  giving exactly three unknown $\alpha_{i}$ function as the following:
  \begin{equation}\label{65.5}
  \boxed{
    \begin{split}
    	 &\alpha_{1}=b \,\frac{ \,G(s)+\,F(s)(s^4-1)}{1-s^4}\\
    	&\alpha_{2}=\,\frac{s+ \,(G(s)-F(s))(s^4+1)}{s^2} \\
    	&\alpha_{3}=b\,\frac{(G(s)+F(s))(s^8-1)+\,s (1-s^6)}{s^2(s^4-1)}    
    \end{split}}
    \end{equation}
 In summary, up to first order in derivative expansion, the drag force exerted on a heavy quark, moving through a general fluid dynamical flow may be given by the following formula:
  \begin{equation}\label{final}
    \begin{split}
     F^{\mu}_{(0)+(1)}(x)&=\frac{\sqrt{\lambda}}{2\pi\, b^2}\,\left(1+\alpha_{1}\,(\tilde{u}.\partial )S_{1}+\,\alpha_{2}\, (\tilde{u}.\partial )S_{2}+\frac{}{} \alpha_{3}\,(u.\partial )S_{1} \right)\,w^{\mu}\\
     &+   \frac{\sqrt{\lambda}}{2\pi} \left(\tilde{\beta}_{1}\,(\tilde{u}^{\mu}(\tilde{u}.\partial)S_{1}+\partial^{\mu}S_{1})+\frac{}{}\tilde{\beta}_{2} \,(\tilde{u}.\partial) w^{\mu} +\, \tilde{\beta}_{3}\, (u.\partial )w^{\mu}   \right)   
     \end{split}
       \end{equation}
where $w^{\mu}=u^{\mu}+\frac{}{}S_{1}\,\tilde{u}^{\mu}$. Here $\sqrt{\lambda}=1/\alpha'$ where the $\lambda$ is the related t'Hooft coupling.

 \subsection{Reproducing the result of arXiv:1202.2737 \cite{Abbasi:2012qz}}
 \label{comparison}
  In \cite{Abbasi:2012qz} we have computed  the drag force exerted on a  transverse quark  with zero rapidity in Bjorken flow.
  In equation (6.15) and (6.16) therein, we have presented the drag force in the quark's RF.
  Although both \cite{Abbasi:2012qz} and the current paper have common origins, there is a subtle difference which 
  distinguishes them from each other. The point is that the procedure used in \cite{Abbasi:2012qz} was 
  clearly dependent on the fluid profile. Due to the analogy between the boundary metric in the RF of transverse quarks
  and metric (7.10) in \cite{Bhattacharyya:2008ji}, we used the  Forced-Fluid/Gravity duality to absorb  the velocity of the Bjorken flow
  into the boundary metric. As a result we dealt only with local temperature as the only fluid variable in \cite{Abbasi:2012qz}. 
  
   Now we want to reproduce 
  those results by use of our general covariant drag formula obtained in previous subsection.
   Before proceeding, let us fix the notation:
   Consider the one dimensional Bjorken flow in the x direction:
   \begin{equation}\label{59}
   u^{\mu}(t,x,y,z)=\frac{1}{\sqrt{t^2-x^2}}(t,x,0,0)\, , \;\;\;\;\;\;\;\; T(t,x,y,z)=\frac{T_0}{(t^2-x^2)^{1/6}}\,.
   \end{equation}
  
  Consider a quark with velocity $\tilde{u}^{\mu}=(\tilde{u}^0,0,0,\tilde{u}^z)$    in the labratory frame.
   In the RF \footnote{Let us take the coordinates in this frame as $(T,X,Y,Z).$} of the quark,
   the fluid velocity and temperature turn out to be
\begin{equation}\label{60}
\begin{split}
 b_{RF}(T,X,Y,Z)&=b_{0} \;(\sqrt{(\tilde{u}^0 T+ \tilde{u}^z Z)^2-X^2})^{1/3}\,,\\
   u^{\mu}_{RF}(T,X,Y,Z)&=(U^0,0,0,U^{Z})\\
  &=\frac{\tilde{u}^0 T+ \tilde{u}^z Z}{\sqrt{(\tilde{u}^0 T+ \tilde{u}^z Z)^2-X^2}}(\tilde{u}^0 ,\frac{X}{\sqrt{(\tilde{u}^0 T+ \tilde{u}^z Z)^2-X^2}},0,-\tilde{u}^z) \,.
  \end{split}
\end{equation}
   In order to use \eqref{45} and \eqref{54.5} we need to compute three independent derivative terms. One can simply show that in the quark  position, $X=0$, 
   \begin{equation}\label{61}
   \partial_{T}u_{RF}^{Z}=\partial_{Z}u_{RF}^{Z}=0\,.
   \end{equation}
   So $\partial_{T}b_{RF}=\dot{b}$ is the only derivative term contributing to the drag in RF.
   Now we can rewrite \eqref{45} and \eqref{46} as following:
      \begin{equation}\label{62}
      F^{Z}_{1}=\frac{1}{2\pi \alpha'}\; \frac{U^{z}}{b_{RF}^2}\;\left[\big(\mathcal{U}-F(\sqrt{U^0})\big)\;\frac{(1+U_{0}^2)}{U^{0}}+\;\frac{\sqrt{U^0}}{U^0} \right] \dot{b}\,.
   \end{equation}
 In order to reproduce the drag force in desired form we first use equations (6.20) in \cite{Abbasi:2012qz} to simplify \eqref{62}:
  \begin{equation}\label{611}
  \frac{1}{U^{0}}\;\dot{b}_{RF}=\frac{U_{0}^{2}-U_{z}^{2}}{U^{0}}\; \dot{b}_{RF}=\;(-U_{0}\;\dot{b}_{RF}+U_{z}\;b_{RF}')\,.
  \end{equation}
To proceed we only need to change $\dot{b}$. By use of \eqref{59} it can be replaced with:
\begin{equation}\label{612}
   \dot{b}=U^0\;\frac{ b}{3T}\,.
\end{equation}
    Plaguing \eqref{611} and \eqref{612} in  \eqref{62}, $F^{Z}_{1}$ will be turned out to be in
     complete agreement with  (6.14) and (6.15) in   \cite{Abbasi:2012qz}.
   In the convention of \cite{Abbasi:2012qz}:
   \begin{equation}\label{63}
   F^{Z}= F^{Z}_{(0)}+ F^{Z}_{(1)}= \frac{1}{2\pi\alpha'}\;\frac{U^Z}{b_{RF}^2}\;(1+\textit{A}\; U_{0}\dot b+\textit{B}\; U_{Z} b'+\textit{C}\; \dfrac{b_{RF}}{T})
\end{equation}      
with
\begin{equation}
\begin{split}
&A=-B=\;-\frac{1+U_0^2}{b}\;\mathcal{U}\,,\\
&C=\frac{1}{3 b U^0}\left(\sqrt{U^{0}}-\frac{}{}(1+U_{0}^2)F(\sqrt{U^{0}})\right)\,.
\end{split}
\end{equation}
Finally, it would be interesting to discus a little bit more about the nature of the correction terms in \eqref{63}.   
As we saw in \eqref{61}, there can not be present any correction terms with derivative of the velocity \textbf{ in RF}.
One  does not have to think that this  is the only representation of derivative terms for Bjorken flow. Fluid dynamics
equations of motion allows us to present derivative terms in some various combinations. For instance the
 second line of \eqref{62} which seems as a $b$ derivative term here, has appeared from $\partial u$ term  in \cite{Abbasi:2012qz}:
\begin{equation}\label{64}
\partial_{\mu}u^{\mu}=\partial_{\mu}(\frac{t}{\sqrt{t^2-x^2}},\frac{x}{\sqrt{t^2-x^2}},0,0)=\frac{1}{\sqrt{t^2-x^2}}=\frac{1}{\tau}\,.
\end{equation} 
Such freedom in presenting the correction terms can be used to exchange the derivative terms in same order with each other.
 As we have shown in \eqref{322}, we have always replaced the divergence of the velocity with some other derivative terms throughout 
 our computations\footnote{In our present case, $u^{\alpha}\partial_{\alpha}u_{z}$  vanishes.}.

 \section{Open questions}
 \label{open}
Our result reported in this paper may be encountered with several follow-up questions. 
We list some of them  below and leave the answers to our future work.

Our main result in this paper is finding the general form of the covariant drag force exerted on a quark
moving through a general fluid flow perturbatively, up to first order in derivative expansion.
As the first extension  one may be interested in  finding the second order derivative corrections.
Basically it would be a solvable problem. To proceed, firstly the most general two derivative 
terms contributing to \eqref{6} have to be determined.
These terms can be generally classified in two sets. The first set
of terms are independent two derivative data, $I_{2}$ data, that have been listed in Table 5
\footnote{Considering the five(four) scalar(vector) equations of motion, one can drop every arbitrary  five-term(four-term) set of scalar(vector) data
to introduce an inequivalent basis of  scalar(vector) terms, namely the $\mathfrak{s}_{i}$($\mathfrak{v}^{\mu}_{i}$) terms.}.
\begin{table}
\renewcommand\arraystretch{1.1}
\centering
\begin{tabular}{| p{1.3 cm} | c | c | c |}
\hline        
	 \raggedright $SO(3)$ & All  data & Equations of motion & $I_{2}$  data\\
\hline
		\hline
	\multirow{7}{*}{Scalars} & 
	$ (\tilde{u}.\partial) (\tilde{u}.\partial) (\tilde{u}.u)$  & 
	\multirow{7}{*}{
		\begin{minipage}{4 cm}
			\begin{flushleft} 
				$(\tilde{u}.\partial)\partial_{\mu}(T^{\mu \nu}\tilde{u}_{\nu}) =0$ \\ 
                                $\;\;u_{\nu}(\tilde{u}.\partial)\partial_{\mu}T^{\mu \nu} =0$ \\
                                $(u.\partial)\partial_{\mu}(T^{\mu \nu}\tilde{u}_{\nu}) =0$\\
                                $\;u_{\nu}(u.\partial)\partial_{\mu}T^{\mu \nu} =0$\\
                                $\;\;\;\;\;\partial_{\mu}\partial_{\nu}T^{\mu\nu}=0$
			\end{flushleft}
		\end{minipage} 
	} &
	\multirow{7}{*}{
		\begin{minipage}{0.5 cm}
			\begin{flushleft} 
				$\mathfrak{s}_1$ \\ 
                                $\mathfrak{s}_2 $ \\  
                                $\mathfrak{s}_3 $   \\  
                                  $\mathfrak{s}_{4}$\\ 
                                   $\mathfrak{s}_{5}$ 
			\end{flushleft}
		\end{minipage} 
	} \\
	&  $u^{\mu} (u.\partial) \partial_{\mu}(\tilde{u}.u)$& & \\
	&$ (u.\partial) (\tilde{u}.\partial) (\tilde{u}.u)$ & & \\
	& $(\tilde{u}.\partial)(\partial .u)$,\;\;\;$(u.\partial)\;(\partial .u)  $& &\\
	&$\partial^{2}(\tilde{u}.u)$ & &\\
	&$(u.\partial) (\tilde{u}.\partial) b$,\;\; $ (\tilde{u}.\partial)^{2} b$ & &\\
	&  $u^{\mu}(u.\partial)\partial_{\nu}b$,\;\;\;\;$\partial^{2}b$ & & \\

\hline
	\multirow{5}{*}{Vectors} & 
	$\tilde{p}^{\mu \alpha} (\tilde{u}.\partial)^{2} u_{\alpha }$,\;$\tilde{p}^{\mu \alpha} \partial_{\alpha}(\tilde{u}.\partial)(\tilde{u}.u)$ & 
\multirow{5}{*}{
		\begin{minipage}{4 cm}
			\begin{flushleft} 
				$(\tilde{u}.\partial)\partial_{\beta}({\tilde{p}^{\mu}}_{\alpha}T^{\alpha \beta}) =0$ \\ 
                                $\tilde{p}^{\mu \alpha} \partial_{\alpha}\partial_{\nu}(T^{\nu \beta} \tilde{u}_{\beta})=0$ \\
                                $\;(u.\partial)\partial_{\alpha}({\tilde{p}^{\mu}}_{\nu}T^{\alpha \nu})=0$\\
                                $\;\;\tilde{p}^{\mu \alpha} u_{\nu} \partial_{\alpha}\partial_{\beta}T^{\nu \beta} =0$
        \end{flushleft}
		\end{minipage} 
	} &	
	\multirow{5}{*}{
		\begin{minipage}{0.5 cm}
			\begin{flushleft} 
				$\mathfrak{v}_1^{\mu} $ \\ 
                                $\mathfrak{v}_2^{\mu} $ \\
                                $\mathfrak{v}_3^{\mu} $\\
                                $\mathfrak{v}_4^{\mu} $\\
                                $\mathfrak{v}_5^{\mu} $
			\end{flushleft}
		\end{minipage} 
	} \\
	& $\tilde{p}^{\mu \alpha}(u.\partial) (\tilde{u}.\partial) u_{\alpha}$,\;\;$\tilde{p}^{\mu \alpha}\partial^{2}u_{\alpha} $ & & \\
	&$\tilde{p}^{\mu \alpha} u^{\beta} (u.\partial) \partial_{\beta} u_{\alpha }$,\;\;$\tilde{p}^{\mu \alpha}\partial_{\alpha}(\partial.u)$  & & \\
	& $\tilde{p}^{\mu \alpha}  (u.\partial)\partial_{\alpha}(\tilde{u}.u)$ & & \\
	& $\tilde{p}^{\mu \alpha}(u.\partial)\partial_{\nu}b$ ,\;$\tilde{p}^{\mu \alpha}(\tilde{u}.\partial)\partial_{\nu}b$  &&\\
\hline
	\end{tabular}
	\caption{$I_{2}$ data.}
\end{table}
The second set,  namely the $I_{1,1}$ data, contains the  products of one derivative terms.
This set of two derivative terms has been organized in table 6.

Accordingly, there are 17 scalar terms and 14 vector terms 
which may contribute to the covariant drag force at second order.
 In order to specify the 31 unknown coefficient functions corresponding to these two derivative terms, we must
 implement the second order analogue of our computations in section\eqref{fluid}
\footnote{Throughout such computations, we will deal with $X_{i}^{(1)}$ function. Using the results obtained in this paper, 
we can  integrate \eqref{15} to find them.}. Will the pure vectorial corrections be absent at this order as well as their 
absence at first order(See \eqref{56}.)? If so, What can be the reason behind it?
 Any answer to this question will be a prediction in the framework of holography, something unpredictable in a general 
 filed theory. This is a specific of the $\mathcal{N}=4$ SYM field theory.

Let us recall that the common point in \cite{Herzog:2006gh}, \cite{Gubser:2006bz} and present work is that in all of them
a quark is restricted to move uniformly through the medium. The problem of shooting a quark 
through the plasma and investigating its energy loss has not  been generally resolved yet and is remained as an open question. Although some 
prescriptions have been suggested to compute the drag force at late times in the plasma \cite{Herzog:2006gh},  there does not exist a clear picture
explaining  the whole-time dynamics. It would be interesting to explore the problem firstly in the equilibriated
plasma and then in the hydrodynamic regime. In the latter case, if analytically possible, some new derivative terms may appear
due to the quark acceleration.

\begin{table}
\renewcommand\arraystretch{1.1}
\centering
\begin{tabular}{| p{1.3 cm} | c |}
\hline
		\hline
	\multirow{2}{*}{Scalars} &$s_{1}^2$,\;\;$s_{2}^2$,\;\;$s_{3}^2$,\;\;$s_{1}s_{2}$,\;\;$s_{2}s_{3}$,\;\;$s_{3}s_{1}$,\;\\
              &$v_{1}^2$,\;\;$v_{2}^2$,\;\;$v_{3}^2$,\;\;$v_{1}.v_{2}$,\;\;$v_{2}.v_{3}$,\;\;$v_{3}.v_{1}$\\
             
	\hline
\hline
	\multirow{3}{*}{Vectors}  & $s_{1}v_{1}^{\mu}$,\;\; $s_{1}v_{2}^{\mu}$,\;\; $s_{1}v_{3}^{\mu}$,\;\;  \\
              & $s_{2}v_{1}^{\mu}$,\;\; $s_{2}v_{2}^{\mu}$,\;\; $s_{2}v_{3}^{\mu}$,\;\;  \\
              & $s_{3}v_{1}^{\mu}$,\;\; $s_{3}v_{2}^{\mu}$,\;\; $s_{3}v_{3}^{\mu}$\;\;  \\ 
\hline
\hline
	\end{tabular}
	\caption{$I_{1,1}$ data}
\end{table}

 In another direction, our covariant formula might be used to study heavy quarks moving through the Bjorken flow in the QGP
 experiment. Having the phenomenological rate of the energy loss of probe quarks, one may proceed to produce the $R_{AA}$
 plot analogue of what has been obtained in \cite{Das:2012jr}. 
Such plot would be the full prediction of AdS/CFT duality up to first order in derivative expansion. 
 We leave more study on the issue to our future work.

\subsection*{Acknowledgements}

We would like to acknowledge useful discussions  with  A. Akhavan.
We wish to express our deep appreciation to H. Arfaei who was generous enough to share his
expertise with us and reading the article through. 

 \section{Note Added in v3: Comparison with Lekaveckas-Rajagopal}
 Shortly after our paper was accepted for publication, a similiar  work by Lekaveckas and Rajagopal appeared on arXiv \cite{Lekaveckas:2013lha}. Their covariant drag formula has some deviations from our covariant result. 
 
 \subsection{Dissagreements}
 As expressed in \cite{Lekaveckas:2013lha}, the position of world-sheet horizon is unaffected by fluid gradiants up to first order (see the paragraph below formula (4.6) in \cite{Lekaveckas:2013lha}). In addition as stated by the authors of \cite{Lekaveckas:2013lha}, there is an extra assumption behind our perturbative structure (see paragraph below formula (2.24) in \cite{Lekaveckas:2013lha}). In the following we disscus about these two comments
 .
 
 Fisrt, in order to investigate whether the position of worldsheet horizon is affected by fluid gradiant, we follow the method of finding a dynamical event horizon developed in \cite{Bhattacharyya:2008xc}. The worldsheet horizon is a null surface on the worldsheet. In the equilibrium,
 this surface can be simply found by this requirement that the vector normal to the surface $S=r-r_H=0$ be null:
 \begin{equation}
 n_\alpha=\partial_{\alpha}S:\,\,\,\,\,\,\,\,\,n_{\alpha}n^{\alpha}=0
 \end{equation}   
 which gives the familiar result $r_H=\frac{\sqrt{u^0}}{b}$. Out of equilibrium however, the equation of surface changes as
 \begin{equation}
 S_{w.h.}(\lambda_{p}(t, \vec{x}))=\,r-r_{H}(\lambda_{p}(t, \vec{x}))=\,0
 \end{equation}
 where again, $\lambda_{p}(t, \vec{x})$ denotes the set of hydro variables.
 Analogous to  what the authors of \cite{Bhattacharyya:2008xc} have done to find the horizon of a black brane, one expects $r_H$ to become a local function of hydro varibles here
 \begin{equation}
 r_{H}(\lambda_{p}(t, \vec{x}))\equiv r_{H}(\lambda_{p}(t, \vec{x}(r_{H},t)).
 \end{equation}
 Since $\vec{x}(r_{H},t)$ denotes the profile of string in the bulk, $r_H$ is only a function of time variable. So we may write:
 \begin{equation}
 n_{\alpha}=\,(n_{r}, n_{t})=\,(1,\,-\,\epsilon\,\frac{\partial r_{H}}{\partial \lambda_{p}}\,\frac{d{\lambda_{p}}}{d t})
 \end{equation} 
 with $\epsilon$ being the book keeping parameter which counts the number of derivatives.
 Applying the null condition, we have
 \begin{equation}\label{7a}
 \begin{split}
 n^{\alpha}n_{\alpha}&=\,g^{\alpha \beta} n_{\alpha}n_{\beta}\\
 &=g^{rr}+2\, \epsilon \, g^{rt} (-\partial_{\lambda\, p}r_{H}\, \dot{\lambda_{p}})+\, \epsilon^2\,g^{tt\,2}\,(\partial_{\lambda\, p}r_{H}\, \dot{\lambda_{p}})^2\\
 &=\frac{1}{g}\, \left(g_{tt}+2\, \epsilon \, g_{rt} (\partial_{\lambda\, p}r_{H}\, \dot{\lambda_{p}})+\, \epsilon^2\frac{}{}g_{rr}^2\,(\partial_{\lambda\, p}r_{H}\, \dot{\lambda_{p}})^2 \right)\,=\,0.
 \end{split}
 \end{equation}
 In order to find the solution of the above equation in a derivative expansion we write:
 \begin{equation}\label{666}
 r^{*}_{H}=\,r^{*}_{(0)H}(\lambda_{p}(\vec{x^*},t))+\,r^{*}_{(1)H}(\lambda_{p}(\vec{x^*},t))
 \end{equation}
 and after substituting this solution into equation (\ref{7a}) we reach to:
 \begin{equation}\label{WH}
 \frac{\partial G_{tt}^{(0)}}{\partial r_{H}}\,r^{*}_{(1)H}=\,
 \,-g_{tt}^{(1)}(r^{*}_{(0)H})-\,\partial_{\lambda\, p}r^{*}_{(0)H}\, \dot{\lambda_{p}}.
 \end{equation}
 If the RHS of (\ref{WH}) vanishes, the position of world-sheet horizon will not change at least to first order in derivatives. By using the following formulas, 
 \begin{eqnarray}
 & g_{tt}^{(1)}=\,G_{tt}^{(1)}+\,2G_{tz}^{(0)}\dot{X}_{Z}^{(0)}\\
 & G_{tt}^{(1)}(r^{*}_{(0)})=\;-\frac{2 F(\sqrt{u^{0}})}{b}\frac{u_{z}^2}{u_{0}}\;\partial_{z} u_{z}-
 \frac{2}{b}\left(u_{z}^2F(\sqrt{u^{0}})-(\sqrt{u^{0}})^3\right)\; \partial_{t} b\\   
 &\dot{X}_{z}^{(0)}(r^{*}_{(0)})=\;\left(\arctan(\sqrt{u^0})-\frac{\pi}{2} \right )\partial_{t}(u_{z} b) +\;\frac{u_{z} \sqrt{u^{0}}}{1+u^{0}} \;\partial_{t}b\\
 &\partial_{\lambda\, p}r^{*}_{(0)H}\,\dot{\lambda_{p}}=\frac{1}{2}\left(\frac{1}{2}\partial_{t}\ln u_{0}-\,\partial_{t} \ln b \right)
 \end{eqnarray}
 it would be a simple calculation to show that actually,
 the latter does not happen.
 Therefore, no matter whether one finds the drag force via the method used in our paper or through the method used in \cite{Lekaveckas:2013lha}, we showed that the position of world-sheet horizon is affected by the fluid gradiant at first order; in contrast to what has been explicitly expressed in \cite{Lekaveckas:2013lha}. More importantly, that the position of world-sheet horizon is affected or unaffected by the fluid gradiants, does not influence on our results. In the next paragraph we will clarify this point.
 
 Second, it has to be emphasized that in both \cite{Abbasi:2012qz} and this paper, we did not make an assumpation as "the effect of  fluid  gradients  could  be  attributed  to  their
 effects on the position of the world-sheet horizon in the dual gravitational description" as stated in \cite{Lekaveckas:2013lha}. Instead, we have explicitly shown that to integrate the equations of motion perturbatively, one could  extend the method that is usually used to compute the drag force in equilibrium\cite{Gubser:2006bz,Herzog:2006gh}. 
 
 Our method is based on this simple fact that both in equilibium and in the situation slightly deviated from equilibrium, the string is trailed all the way in the bulk from boundary to the horizon of AdS black brane; let us remind that this is the common point between 
 \cite{Lekaveckas:2013lha} and our work. As a result, the square root in (\ref{15}) must be well-defined. In equilibrium, this condition fixes the value of $\pi$ at the simple root of numerator, namely at $r=\frac{\sqrt{u^0}}{b}$. Out of local equilibrium however, the coefficients in equation $B_m^2-A_m C_m=0$
 are no longer constant; they are local function of hydro variables with first order derivative corrections. Consequently, $r=\frac{\sqrt{u^0}}{b}$ is now the solution for  $B_m^2-A_m C_m=0$ only up to zero order in derivatives. We have explicitly shown that $r^*=r^*_{(0)}+r^*_{(1)}$ solves this equation up to first order where $r^*_{(1)}$ is given by (\ref{41}). It is worth noting that at zero order, $r^*$ is preciesly the position of world-sheet horizon. Comparing (\ref{41}) with (\ref{666}), it is straightforward to check that at first order
 \begin{equation}
 r^*_{(1)}\ne r^*_{(1)H}.
 \end{equation} 
 It can be clearly seen that we did not assumed that the root of numerator would be corrected, in fact it was the derivative expansion that forced $r^*$ to get a first order correction as well.
 
 Correspondingly, the corrected root of $B_m^2-A_m C_m=0$ ,namely $r^*$, specifies the value of $\Pi$ at the same point. Therefore, without any assumptions, some new contributions appear in the expression of drag force. 
 
 \subsection{Agreements: Drag Forec in Bjorken Flow}
 As one of the applications of covariant drag force formula, in \cite{Lekaveckas:2013lha} the drag force exerted on a quark moving in Bjorken flow has been computed. In what follows we will show that although in a general flow the drag formula given in \cite{Lekaveckas:2013lha} dissagrees with that of us given in current paper, in the Bjorken flow specially, the result of \cite{Lekaveckas:2013lha} is exactly the same as what our formula gives.

 The velocity of Bjorken flow in the laborotary frame may be given by (see \cite{Abbasi:2012qz} for more details):
 \begin{equation}
 u^{\mu}=(\frac{t}{\tau},0,0,\frac{z}{\tau})=\,\gamma(1,0,0,\frac{z}{t})  \,,\,\,\,\,\,\,\,\,\,\,\,\tau=\sqrt{t^2-z^2}.
 \end{equation} 
The inverseof temperaturte is given by:
  \begin{equation}
  b(t, \vec{x})=b(\tau)\propto \frac{1}{\tau^{1/3}}
  \end{equation}
 Suppose a quark produced at $t=0$ and $\vec{x}=0$. In the $z$ direction, the quark moves with the same velocity of flow. In the transverse directions however, it may be pulled with constant velocities $\tilde{v}_x$ and $\tilde{v}_y$. We may write the four-velocity of quark as
  \begin{equation}
  \tilde{u}^{\mu}=\tilde{\gamma}\,\left(1,\tilde{v}_{x},\tilde{v}_{y},\tilde{v}_{z}=\frac{z}{t}\right)
  \end{equation}
 
 In order to utilize (\ref{final}) for our present case, we first simplify the 
 expressions of $S_1$, $S_2$ and $w^{\mu}$
  \begin{equation}\begin{split}
  S_{1}= \,& -\tilde{\gamma}\sqrt{1-z^2/t^2}=\,-\frac{\tilde{\gamma}}{\gamma}\\
  S_{2}=& \,\,b(\tau)\\
  w^{\mu}= & \,\gamma\left(1-\frac{\tilde{\gamma}^2}{\gamma^2}\,\,,\,\, -\frac{\tilde{\gamma}^2}{\gamma^2} \tilde{v}_{x}\,\,,\,\, -\frac{\tilde{\gamma}^2}{\gamma^2}\tilde{v}_{y}\,\,,\,\,\frac{z}{t}-\frac{\tilde{\gamma}^2}{\gamma^2}\tilde{v}_{z}\right).
  \end{split}
  \end{equation}
  Since the above objects are functions of $z$ and $t$. only four following partial derivatives may appear in our computations to first order:
   \begin{equation}
   \partial_{t}u^{0}=-\frac{\tilde{\gamma}^2 \tilde{v}_{z}^2}{\tau}\,,\,\,\,\,\,\,\partial_{t}u^z=-\partial_{z}u^{0}=-\frac{\tilde{\gamma}^2 \tilde{v}_{z}}{\tau}\,,\,\,\,\,\,\,\partial_{t}u^{z}=-\frac{\tilde{\gamma}^2 }{\tau}.
   \end{equation}
  Using the abov ederivatives one can show that the only non-vanishing derivative contribution in $F^{\mu}$ comes from the derivative of $S_2$:
    \begin{equation}
  F^{\mu}_{(0)+(1)}(x)=\frac{\sqrt{\lambda}}{2\pi\, S_{2}^2}\,\left(1+\frac{}{}\alpha_{2}(S_1)\, (\tilde{u}.\partial )S_{2}\right)\,w^{\mu}
  \end{equation}
  Simplifying the derivative term:  
  \begin{equation*}
  \begin{split}
  (\tilde{u}.\partial )\,S_{2}\,&=\,\tilde{u}^{0}\partial_{t}S_{2}\,+\,\tilde{u}^{z}\partial_{z} S_{2}\\
  &=\,\tilde{\gamma}\, b'\, \frac{t}{\tau}\,+\,\tilde{\gamma}\,
  \tilde{v}_{z}\, b'\, \frac{-z}{\tau}\,=\,b'\,\frac{\gamma}{\tilde{\gamma}}\,=\,\frac{b(\tau)}{3 \tau}
  \end{split}
  \end{equation*}
  yields:
   \begin{equation}
   \boxed{
   \vec{f}(\tau)=\,-\frac{\sqrt{\lambda}}{2\pi\, b(\tau)^2}\frac{\tilde{\gamma}}{\gamma}\,\left(1+\alpha_{2}\left(\frac{\tilde{\gamma}}{\gamma}\right)\frac{b(\tau)}{3 \tau} \right)\, 
   \left( 
   {\begin{array}{c}
   	\tilde{v}_{x}\\
   	\tilde{v}_{y}\\
   	\gamma^2 \tilde{v}_{z}\tilde{v}_{\perp}^2\\ 
   	\end{array} } 
   \right)}
   \end{equation}\label{finalBjorken}
   where $F^{\mu}=\tilde{\gamma} f^{\mu}$ and $\tilde{v}_{\perp}^2=\tilde{v}_{x}^2+\tilde{v}_{y}^2 $. The coefficient function $\alpha_2$ can be read from (\ref{65.5}) as the following:
   \begin{equation}
   c(\gamma)=\,\sqrt{\gamma}+(1+\gamma^2)\left(
   \frac{\pi}{2}-\arctan(\gamma)-F(\gamma) \right).
   \end{equation}
\textbf{  As we explicitly showed, the drag force exerted on a quark moving through the Bjorken flow, namely (\ref{finalBjorken}), is exactly the same as formula (4.5)
  in \cite{Lekaveckas:2013lha}.}\footnote{Note the authors of \cite{Lekaveckas:2013lha} computed the drag force with an extra minus sign as the force needed to pull the quark uniformly in the flow.}

Let us recall that the results reported in \cite{Lekaveckas:2013lha} and current paper dissagree with each other for a general fluid dynamical flow. One may be interested to understand why two different formulas give a common result in Bjorken flow. The things which distinguish the formula (3.35) in \cite{Lekaveckas:2013lha} from the formula \ref{final} in this paper are the coefficients of terms with the velocity gradiants. In Bjorken flow however, these terms all vanish and consequently the formula (4.5) in \cite{Lekaveckas:2013lha} coinsides with (\ref{finalBjorken})
  
  It would be also interesting to investigate why in a general fluid dynamical flow, the result reported in \cite{Lekaveckas:2013lha} disagrees with our  drag formula.

   \appendix
\section{Appendices}
\label{Appendix}

\subsection{Lagrangian of the string}
\label{lagrang}
According to our gauge choice, 
\begin{equation}
\mathcal{L}=\sqrt{g_{rt}^2-g_{rr}g_{tt}}
\end{equation}
with
\begin{equation}\begin{split}
&g_{rt}=\;G_{tt}+\;2\;G_{ti} \dot{X}_{i}+\;G_{i j} \dot{X}_{i} \dot{X}_{j}\\
& g_{rr}=G_{rr}+\;2\;G_{ri}X^{'}_{i}+\;G_{ij} X^{'}_{i}X^{'}_{j}\\
&g_{tt}=\;G_{rt}+\;G_{ri} \dot{X}_{i}+\; G_{ti}X^{'}_{i}+\;G_{ij}\dot{X}_{i}X^{'}_{j}
\end{split}
\end{equation}
where for sake of brevity, we have dropped the argument of derivative terms.

\subsection{Coefficient functions in the Lagrangian \eqref{14}}
\label{coef}
\begin{equation}\begin{split}
 A_{m}=&\;G_{tm}^2-\;G_{mm} G_{tt}+\;2\;(G_{tm} G_{km} - \;G_{mm} G_{tk} )\dot{X}_{k}\\
 &+\;(G_{im}G_{km} -\; G_{mm} G_{ik})\dot{X}_{i}\dot{X}_{k}\end{split}
 \end{equation}
 \begin{equation}\begin{split}
 B_{m}=&\;G_{rt}G_{tm}-\;G_{rm}G_{tt}+\;(G_{t\tilde{i}}G_{tm}-G_{\tilde{i}m}G_{tt})X^{'}_{\tilde{i}}+\;(G_{rt}G_{im}+G_{ri}G_{tm}-2G_{rm}G_{ti})\dot{X}_{i}\\
& +\;(G_{ri}G_{jm}-G_{rm}G_{ij})\dot{X}_{i}\dot{X}_{j}+\;(G_{t\tilde{i}}G_{jm}+G_{tm}G_{\tilde{i}j}-2G_{\tilde{i}m}G_{tj})X^{'}_{\tilde{i}}\dot{X}_{j}\\
&+ \;(G_{\tilde{i}j}G_{km}-\;G_{\tilde{i}m}G_{jk})\dot{X}_{j}\dot{X}_{k}X^{'}_{\tilde{i}}
\end{split}
\end{equation}
\begin{equation}\begin{split}
C_{m}=&G_{rt}^2-\;G_{rr}G_{tt}+\;2(G_{rt}G_{ri}-\;G_{rr}G_{ti})\dot{X}_{i}+\;2(G_{rt}G_{t\tilde{i}}-G_{r\tilde{i}}G_{tt})X^{'}_{\tilde{i}}\\
&+\;(G_{\tilde{i}t}G_{\tilde{j}t}-\;G_{tt}G_{\tilde{i}\tilde{j}})X^{'}_{\tilde{i}}X^{'}_{\tilde{j}}+\;(G_{ri}G_{rj}-\;G_{rr}G_{ij})\dot{X}_{i}\dot{X}_{j}\\
&+\;2(G_{rt}G_{i\tilde{j}}+\;G_{ri}G_{t\tilde{j}}-\;2G_{ti}G_{r\tilde{j}})\dot{X}_{i}X^{'}_{\tilde{j}}\\
&+\;2(G_{i\tilde{j}}G_{t\tilde{k}}-\;G_{ti}G_{\tilde{j}\tilde{k}})\dot{X}_{i}X^{'}_{\tilde{j}}X^{'}_{\tilde{k}}
+\;(G_{i\tilde{k}}G_{j\tilde{k}}-G_{ij}G_{\tilde{k}\tilde{l}})\dot{X}_{i}\dot{X}_{j}X^{'}_{\tilde{k}}X^{'}_{\tilde{l}}
\end{split}
\end{equation}

\subsection{Quark in a globally boosted thermal plasma}
\label{glob}
Since the thermal state and its dual gravity are time independent, the $X_{i}(r,t)$ in \eqref{9}
will be just functions of $r$ coordinate. Let us start by the boost direction, considering $X_{m}(r)$ for the case of $m=z$.
Using \eqref{coef} it is straightforward to obtain
\begin{equation}\label{18}
X^{'}_{z}(r)=\;\frac{r^2 u_{z}}{r^4f(b r)}\pm\;\frac{\pi_{z}(r)}{r^4f(b r)}\;\sqrt{\frac{r^4 u_{z}^2-r^4 f(b r) u_{0}^2}{\pi_{z}(r)^2-r^4f(b r)}}\,.
\end{equation}
The numerator  of the square root in the second term has a simple root at $r^{*}=\sqrt{u^{0}}/b$.
In order to string to be trailed from boundary to the horizon, the denominator must vanish at the same point.
This requirement fixes the value of $\pi_{z}$ at $r^{*}$:
\begin{equation}\label{19}
\pi_{z}(r^{*})=\;- \frac{u_{z}}{b^2}\,.
\end{equation}
Integrating \eqref{10}  gives exactly the same constant value for $\pi_{z}(r)$ at any other point on the string. 
So, the z component of the covariant drag force and the z-component of the embedding are determined 
as we have demonstrated in table 3.

In the case of transverse directions we again  utilize \eqref{coef} to rewire  equation \eqref{15} for $m=\;x,y$ directions:
 \begin{equation}\label{20}
 X^{'}_{x,y}(r)=\;\pm\;\frac{\pi_{x,y}(r)}{r^4f(b r)}\;\sqrt{\frac{-(r^4 f(b r) -u_{z}^2/b^4)}{\pi_{x,y}(r)^2-(r^4 f(b r) -u_{z}^2/b^4)}}\,.
 \end{equation}
  Just like in the boost direction the numerator  vanishes at $r^{*}$. But in contrast to  \eqref{18} there is no similar way here to remove the 
 sign change of the numerator. The only possibility  is to force $\pi_{x,y}(r)$ to vanish everywhere on the world-sheet.
 As a result, there would be no drag force in these directions (see \eqref{11},\eqref{12}). In addition \eqref{20} directly gives the  $x$ and $y$ components
 of the embedding as it i  shown in table 3.

\subsection{Intermediate steps of  computation in $m=z$ direction}
\label{middle}

\subsubsection{The coefficient functions}
\label{first order coef}
Notice that the first therm in each of following three expressions is a ''local" zero order term which according
to convention below \eqref{29'}, has a first derivative Taylor part contributing to first order corrections in the patch. However
we have argued that in order to determine $r^{*}$ through computing $B_{z}^2-A_{z}C_{z}$, we only need to write 
down these coefficients at $(t,\overrightarrow{x})=\;(0,0)$.
\begin{equation}\label{Ap6}
A_{z}=\;r^4 f(b r)+\;\frac{2u_{0}u_{z}}{r^2 f(b r)}G_{tz}^{(1)}-r^2(\frac{u_{0}^2}{r^4 f(b r)}-1)G_{zz}^{(1)}-r^2(\frac{u_{z}^2}{r^4 f(b r)}+1)G_{tt}^{(1)}
\end{equation}
\begin{equation}\label{Ap7}
B_{z}=\;-r^2 u_{z}+\;2(u_{z}G_{tt}^{(1)}-u_{0}G_{tz}^{(1)})-u_{0} r^2 \dot{X}_{z}^{(0)}
\end{equation}
\begin{equation}\label{Ap8}
C_{z}=\;u_{0}^2+\;2u_{0}u_{z} \dot{X}_{z}^{(0)}
\end{equation}
As we have seen in the text, the correction part of these coefficients are always evaluated at $r=r^{*}_{(0)}$. In the following
we only give those corrections  that contribute to the next expressions, evaluated at $r=r^{*}_{(0)}$:

  \begin{equation}\begin{split}\label{}
 &G_{tz}^{(1)}(r^{*}_{(0)})=\;\frac{1}{b}\sqrt{u^{0}}\;\partial_{t} u_{z}-\;\frac{1}{b}(2u_{0}^2F(\sqrt{u^{0}})-\frac{u_{z}}{\sqrt{u^{0}}})\;\partial_{z} u_{z}-\;\frac{2u_{z}}{b^2}(u_{0}F(\sqrt{u^{0}})+\sqrt{u^{0}})\;\partial_{t} b\\
  &G_{tt}^{(1)}(r^{*}_{(0)})=\;-\frac{2 u_{z}^2}{u_{0}\;b}\;\partial_{z} u_{z}-
 \frac{2u_{z}}{b^2}(u_{z}F(\sqrt{u^{0}})-\frac{(\sqrt{u^{0}})^3}{u_{z}})\; \partial_{t} b
\end{split}
\end{equation}
Notice that since in \eqref{Ap6} the term containing $G_{zz}^{(1)}$  vanishes at $r^{*}_{(0)}$, $G_{zz}^{(1)}$ does not arise 
in our computations and we do not need to know its expression.

Another useful expression is:
 \begin{equation}
 \dot{X}_{z}^{(0)}(r^{*}_{(0)})=\;\left(\arctan(\sqrt{u^0})-\frac{\pi}{2} \right )\partial_{t}(u_{z} b) +\;\frac{u_{z} r b}{1+r^2 b^2} \;\partial_{t}b
 \end{equation}
 %
\subsubsection{EoM in detail to first order}
\label{detailed}
%
\begin{equation}\begin{split}
f_{z}=\;\frac{\partial \mathcal{L}}{\partial z}&=\;\epsilon \;\frac{-1}{2\pi \alpha'}\;\; \frac{1}{2}  \;\partial_{z} \left(A^{(0)}_{z}\; (X_{z}^{(0)})^{'2}+ \frac{}{} 2B^{(0)}_{z}\;(X_{z}^{(0)})'+\;C^{(0)}_{z} \right) + O(\epsilon^2)\\
&= \;0+\;O(\epsilon^2)
\end{split}
\end{equation}
\begin{equation}\begin{split}
\Pi^{t}_{z}(r,t)=\;\frac{\partial \mathcal{L}}{\partial \dot{X}_{z}}&= \; \frac{-1}{2\pi \alpha'}\;\left(u_{0}^2-\frac{}{}2u_{z}r^2\;(X_{z}^{(0)})'+\;r^4 f(r b) \; (X_{z}^{(0)})^{'2}\right)+\;O(\epsilon)\\
&=\;\frac{1}{1+r^2 b^2}\;u_{0}u_{z}+\; O(\epsilon)
\end{split}
\end{equation}
Notice that in the equation above, we have written only zero order part of the $\Pi^{t}_{z}(r,t)$. The reason for this is that
EoM contains time derivative of $\Pi^{t}_{z}(r,t)$ which is one upper order  than $\Pi^{t}_{z}(r,t)$ itself.

 In the expression below we give the manipulated integral term of \eqref{44}:
\begin{equation}\begin{split}
(-2\pi \alpha')\int_{r^{*}(t)}^{\infty}(f_{z}-\; \partial_{t}\Pi^{t}_{z})\;d \mathbf{r}\;=&\;-\;\epsilon\;\;\mathcal{U}\;\frac{1}{b}\;\left(u_{0}+\;\frac{u_{z}^2}{u_{0}} \right)\;\partial_{t} u_{z}\\
&+\;\epsilon\;\;\mathcal{V}\; \frac{1}{b^2}\;u_{0} u_{z}\;\partial_{t} b \;+\;O(\epsilon^2)
\end{split}
\end{equation}
with
\begin{equation}
\begin{split}
&\mathcal{U}=\;\;\frac{\pi}{2}-\arctan(\sqrt{u^{0}})\,, \\
 & \mathcal{V}=\; -\;\mathcal {U}+\;\frac{\sqrt{u^{0}}}{1+u^{0}}\,.
\end{split}
\end{equation}
%
\subsection{Intermediate steps of  computation in $m=\{x,y\}=T$ directions}
\label{middle T}

%
\begin{equation}\label{A15}
A_{T}=\;-r^4(\frac{u_{0}^2}{r^4 b^4}-1)-\;r^2 \left(G_{tt}^{(1)}+\;(\frac{u_{0}^2}{r^4 b^4}-1)\;G_{TT}^{(1)} \right)-\;2 r^4\;\frac{u_{0}u_{z}}{r^4 b^4} \dot{X}_{z}^{(0)}
\end{equation}   
  \begin{equation}
  B_{T}=\;-u_{0}\;G_{tT}^{(1)}+\;r^2 \left(\frac{u_{0}u_{z}}{r^4 b^4}\;G_{tT}^{(1)}-\;(\frac{u_{0}^2}{r^4 b^4}-1)\;G_{zT}^{(1)} \right)\;\dot{X}_{z}^{(0)}
  \end{equation} 
\begin{equation}
\begin{split}
C_{T}&=\;1+\;2\;(u_{z} G_{tt}^{(1)}-\;u_{0} G_{tz}^{(1)})\;(X_{z}^{(0)})'-\;2 r^2 u_{z}\;(X_{z}^{(1)})'\\
&+\;r^2\left( 2(\frac{u_{0}u_{z}}{r^4 b^4}) G_{tz}^{(1)}-\;(\frac{u_{z}^2}{r^4 b^4}+1)G_{tt}^{(1)}+\;(\frac{u_{0}^2}{r^4 b^4}-1)G_{zz}^{(1)} \right)\;(X_{z}^{(0)})^{'2}\\
&+\; 2\; f(b r)\; (X_{z}^{(0)})' (X_{z}^{(1)})'+\;2\;u_{0} u_{z}(1-\;(X_{z}^{(0)})') \;\dot{X}_{z}^{(0)}
\end{split}
\end{equation}  
 Notice that In \eqref{48} we have written the zero order part of these coefficients.
  
 Recall that for $T$ directions, Lagrangian is given by:
\begin{equation}
\mathcal{L}=\;\sqrt{A_{T} X_{T}^{'2}+\;2B_{T} X_{T}^{'}+\;C_{T}}.
\end{equation}   
 Considering the coefficients given in above and that the $X_{T}$ appears at first order in gradient expansion,
 we understand that  $A_{T} X_{T}^{'2}+\;2B_{T} X_{T}^{'}$ can not contribute to $f_{T}$  up to first order.  Notice that derivative with 
 respect to $T$ increases the order of  every term present in $\mathcal{L}$ by one. So the  only term whose $T$ derivative 
 might survive to first order is the zero order part of $C$ which is itself a constant independent of all coordinates. Consequently,
 \begin{equation}
 f_{T}=\;0+\;O(\epsilon^2)
  \end{equation}

The other necessary quantity is $\Pi^{t}_{T}=\;\partial \mathcal{L}/ \partial \dot{X}_{T}$. As it was noted above
 there do not exist any $T$ derivative terms in $\mathcal{L}$ up to first order. It means that:
 \begin{equation}
 \Pi^{t}_{T}=\;0+\;O(\epsilon^2)
  \end{equation}

\bibliographystyle{utphys}

\begin{thebibliography}{10}


\bibitem{Shuryak:2003xe}
  E.~Shuryak,
  ``Why does the quark gluon plasma at RHIC behave as a nearly ideal fluid?,''
  [hep-ph/0312227].


\bibitem{Maldacena:1997re}
  J.~M.~Maldacena,
 ``The Large N limit of superconformal field theories and supergravity,''  
    [hep-th/9711200].


\bibitem{Witten:1998qj}
  E.~Witten,
  ``Anti-de Sitter space and holography,'' 
  [hep-th/9802150].  








\bibitem{Herzog:2006gh}
  C.~P.~Herzog, A.~Karch, P.~Kovtun, C.~Kozcaz and L.~G.~Yaffe,
  ``Energy loss of a heavy quark moving through N=4 supersymmetric Yang-Mills plasma,''
  [hep-th/0605158].

 \bibitem{Gubser:2006bz}
  S.~S.~Gubser,
  ``Drag force in AdS/CFT,''
  [hep-th/0605182].

\bibitem{Horowitz:2007su}
  W.~A.~Horowitz and M.~Gyulassy,
  ``Heavy quark jet tomography of Pb + Pb at LHC: AdS/CFT drag or pQCD energy loss?,''
  [arXiv:0706.2336 [nucl-th]].


\bibitem{Kolb:2003dz}
  P.~F.~Kolb and U.~W.~Heinz,
 ``Hydrodynamic description of ultrarelativistic heavy ion collisions,''  
   [nucl-th/0305084].  


\bibitem{Bjorken:1982qr}
  J.~D.~Bjorken,
  ``Highly Relativistic Nucleus-Nucleus Collisions: The Central Rapidity Region,''
  Phys.\ Rev.\ D {\bf 27}, 140 (1983).

\bibitem{Bhattacharyya:2008jc}
  S.~Bhattacharyya, V.~E. Hubeny, S.~Minwalla and M.~Rangamani,
  ``Nonlinear Fluid Dynamics from Gravity,''
  [arXiv:0712.2456 [hep-th]].

\bibitem{Abbasi:2012qz}
  N.~Abbasi and A.~Davody,
  ``Moving Quark in a Viscous Fluid,''
  JHEP {\bf 1206} (2012) 065
  [arXiv:1202.2737 [hep-th]].
  
\bibitem{Chesler:2007sv}
  P.~M.~Chesler and L.~G.~Yaffe,
  ``The Stress-energy tensor of a quark moving through a strongly-coupled N=4 supersymmetric Yang-Mills plasma: Comparing hydrodynamics and AdS/CFT,''
  [arXiv:0712.0050 [hep-th]].
  
\bibitem{Bhattacharya:2011tra}
  J.~Bhattacharya, S.~Bhattacharyya, S.~Minwalla and A.~Yarom,
  ``A Theory of first order dissipative superfluid dynamics,''
  arXiv:1105.3733 [hep-th].

\bibitem{Bhattacharyya:2008xc}
  S.~Bhattacharyya, V.~E.~Hubeny, R.~Loganayagam, G.~Mandal, S.~Minwalla, T.~Morita, M.~Rangamani and H.~S.~Reall,
  ``Local Fluid Dynamical Entropy from Gravity,''
  [arXiv:0803.2526 [hep-th]].


\bibitem{Bhattacharyya:2008ji}
  S.~Bhattacharyya, R.~Loganayagam, S.~Minwalla, S.~Nampuri, S.~P.~Trivedi and S.~R.~Wadia,
  ``Forced Fluid Dynamics from Gravity,''
  [arXiv:0806.0006 [hep-th]].


\bibitem{Das:2012jr}
  S.~K.~Das and A.~Davody,
  ``A non-conformal holographic model for D-meson suppression at LHC energy,''
  arXiv:1211.2925 [hep-ph].



\bibitem{Lekaveckas:2013lha}
M.~Lekaveckas and K.~Rajagopal,
JHEP {\bf 1402} (2014) 068
[arXiv:1311.5577 [hep-th]].













\end{thebibliography}

\providecommand{\href}[2]{#2}\begingroup\raggedright\endgroup

\end{document}